\documentclass[11pt]{article}
\usepackage{epsfig} 
\newcommand{\sect}[1]{ \section{#1} \setcounter{equation}{0} }

\newcommand{\pslash}{p \! \! \! /} 
\newcommand{\qslash}{q \! \! \! /} 
\newcommand{\rslash}{r \! \! \! /}

\newcommand{\QEDs}{\mbox{\footnotesize{QED}}} 
\newcommand{\Nc}{N_{\!c}}
\newcommand{\Nf}{N_{\!f}}

\begin{document}
\title{Symmetric point $4$-point functions at one loop in QCD}
\author{J.A. Gracey, \\ Theoretical Physics Division, \\ 
Department of Mathematical Sciences, \\ University of Liverpool, \\ P.O. Box 
147, \\ Liverpool, \\ L69 3BX, \\ United Kingdom.} 
\date{}
\maketitle 

\vspace{5cm} 
\noindent 
{\bf Abstract.} We evaluate the quartic ghost and quark Green's functions as
well as the gluon-ghost, gluon-quark and ghost-quark $4$-point functions of
Quantum Chromodynamics at one loop at the fully symmetric point in a linear
covariant gauge. Similar expressions for the analogous Green's functions in 
Quantum Electrodynamics are also provided. 

\vspace{-15cm}
\hspace{13cm}
{\bf LTH 1123}

\newpage 

\sect{Introduction.}

Quantum Chromodynamics (QCD) has served as the central quantum field theory
which describes the force between the partons of the nucleons. By partons we 
mean the quarks which are bound together by the force carrying quantum which is
the gluon. The theory has been paramount in describing physics at the high
energy scales used in particle accelerators such as the Large Hadron Collider.
What is less clear is the connection between the QCD field theory at low
energy and the properties of hadrons. The latter are the observed states of
nature and if QCD is to explain these particles at a fundamental level then the
field theory needs to be solved at low energy. This is not a straightforward
task. At high energy QCD is asymptotically free, \cite{1,2}, in that the
strength of the coupling constant decreases with the increase in energy. Thus
one can apply perturbative techniques to carry out highly precise computations.
For example, such multiloop methods have recently been employed to determine 
the $\beta$-function of QCD to {\em five} loops in \cite{3} which was
subsequently verified in \cite{4}. As the energy scale decreases the strength 
of the coupling constant increases and therefore the perturbative approximation
ceases to be of use for studying infrared properties of QCD. For instance, one 
would have to have a large number of terms in a series which would then have to
be summed to get anywhere near meaningful results at lower energies. This is of 
course if one overlooks the fact that purely non-perturbative phenomena could 
become relevant at lower energies. By non-perturbative we mean contributions 
which are non-analytic functions of the coupling constant and therefore not 
accessible from perturbation theory. These issues aside the major gap in our 
understanding is the actual mechanism by which quarks condense to form hadrons 
and why quarks and gluons are confined and not observed as free entities in 
nature. To tackle these problems in a theoretical framework requires techniques
which are valid beyond the perturbative regime. The two dominant methods are 
lattice gauge theory and Schwinger-Dyson equations. The former is a numerically
intense approach while the latter extends beyond perturbation theory by 
systematically solving all the $n$-point Green's functions. As there are an 
infinite number of such equations at some point an approximation has to be made
so that they truncate to a finite set of equations which is manageable.

Much progress has been made in QCD using Schwinger-Dyson equations over the 
last forty years or so with the focus being on the $2$- and $3$-point 
functions. One of the truncations which is ordinarily made in the
Schwinger-Dyson approach is to neglect contributions from $4$-point vertex
functions or $4$-point Green's function kernels. This has turned out in general
to be a reasonable assumption. However, with the improvement in computational
tools to provide more precise insight into the infrared behaviour of QCD it is
now the situation where the neglect of $4$-point vertex functions needs to be
re-examined. Indeed in the Schwinger-Dyson study of \cite{5} the effect of 
omitting the quartic gluon vertex function has been numerically quantified. 
More recently there has been interest in investigating the unquenching effects
of the gluonic $3$-point function through what is termed the swordfish diagram
which involves the quark-gluon $4$-point function, \cite{6}. The model for that
Green's function in the Schwinger-Dyson context grew out of ideas developed in 
\cite{7}. While such an approach could in some sense be viewed as a first 
analysis the model suffers from lack of input from explicit field theory 
calculations. For instance, the solution of any set of Schwinger-Dyson 
equations has to agree with high energy perturbative results of the same 
Green's function. For $2$- and $3$-point functions this has been clearly tested
at length and verified over many years. For $4$-point functions the situation 
is less clear. This is mainly because the full off-shell $4$-point functions 
are not known even at one loop in QCD. What is available is the gluon $4$-point
function at one loop at the fully symmetric subtraction point. The original one
loop analysis was carried out in \cite{8}. Although that concentrated on the 
issue of renormalization schemes a more recent analysis was provided in 
\cite{9}  which recorded the full $4$-point function at the symmetric point 
including all possible Lorentz tensor channels. This at least gives a benchmark
contact point for comparison with a Schwinger-Dyson analysis of the same 
function. 

What is lacking now is the same information for the quark-gluon $4$-point 
function, for instance, as it is now becoming relevant for the complete
$3$-point vertex function Schwinger-Dyson construction. Therefore, it is the 
purpose of this article to compile that Green's function at the fully symmetric
point in QCD at one loop similar to the gluon $4$-point function of \cite{8,9}.
In addition we will carry out the same computation for the other $4$-point 
functions which are the gluon-ghost and ghost-quark $4$-point functions as well
as the quartic ghost and quark Green's functions. While these latter Green's 
functions may not be immediately relevant to the discussion of the 
approximation made in \cite{6,7} for the gluonic $3$-point function they will 
be relevant for other $3$-point vertex functions as well as Schwinger-Dyson 
studies of $4$-point functions. Indeed there has been progress in the latter 
respect in recent years with the study of the quartic gluon and quartic quark 
Green's functions. For instance, a non-exhaustive set of articles on these are 
\cite{10,11,12,13,14,15} and \cite{16,17} respectively. At this point we note
that all our computations will be carried out for the canonical linear 
covariant gauge. In this gauge the five QCD Green's functions considered here 
have no tree term in contrast to other gauges. For instance, in the nonlinear 
Curci-Ferrari gauge, \cite{18}, there is a quartic ghost interaction in the 
Lagrangian. While our main focus is in the linear covariant gauge, as this is 
the one relevant for Schwinger-Dyson analyses, the tools provided here can in 
principle be applied to the Curci-Ferrari gauge as well as other gauges. 
Equally as the $4$-point functions can be studied in Quantum Electrodynamics 
(QED) we will provide the same information for the quartic photon and electron 
Green's functions as well as the photon-electron $4$-point function. This will
be a separate evaluation since our QCD computations, like \cite{9}, will be 
performed purely in the $SU(\Nc)$ colour group, where $\Nc$ is the number of 
colours, for which the abelian limit cannot be taken straightforwardly in a 
$4$-point analysis.

The article is organized as follows. The formalism, notation and technical
details of how the computations were carried out are discussed in the next 
section. The subsequent sections record respectively the results for the five
Green's functions of QCD and the three of QED. We provide conclusions in
section $5$. An appendix records the various Lorentz tensor bases which we used
to decompose the various Green's functions into. 

\sect{Formalism.}

We begin our analysis by discussing the general formalism used to evaluate the
Green's functions we are interested in at the fully symmetric point. This will
be based for the most part on the approach used in \cite{9} to determine the
gluon $4$-point function which in the notation we will use was $\left\langle 
A^a_\mu(p) A^b_\nu(q) A^c_\sigma(r) A^d_\rho(-p-q-r) \right\rangle$. Our  
convention will be that the final field of the Green's function is the one
where the conservation of energy-momentum is implemented as $p$, $q$ and $r$ 
are the external momentum. The restriction of these momenta to the symmetric 
point is defined by, \cite{8}, 
\begin{equation}
p^2 ~=~ q^2 ~=~ r^2 ~=~ -~ \mu^2 ~~,~~
pq ~=~ pr ~=~ qr ~=~ \frac{1}{3} \mu^2
\label{symmpt}
\end{equation}
which implies that the Mandelstam variables given by 
\begin{equation}
s ~=~ \frac{1}{2} ( p + q )^2 ~~,~~ t ~=~ \frac{1}{2} ( q + r )^2 ~~,~~
u ~=~ \frac{1}{2} ( p + r )^2 
\end{equation}
are equivalent since 
\begin{equation}
s ~=~ t ~=~ u ~=~ -~ \frac{4}{3} \mu^2 
\end{equation}
where $\mu$ is an arbitrary mass scale. As the gluon and quark fields carry
Lorentz and spinor indices respectively and these fields as well as the 
Faddeev-Popov ghost are labelled with colour group indices the decomposition of
each Green's function is not straightforward. Therefore we need a systematic 
way of extracting the structure of the Green's functions which accommodates 
this complication and allows us to perform the one loop evaluation. Before
discussing this we need to record explicitly the setup of the various Green's 
functions in the form in which they were computed which are 
\begin{eqnarray}
\left. \left\langle \bar{c}^a(p) c^b(q) A^c_\sigma(r) A^d_\rho(-p-q-r)
\right\rangle \right|_{\mbox{\footnotesize{symm}}} &=&
\left. \Sigma^{abcd}_{Ac\,\sigma \rho}(p,q,r)
\right|_{\mbox{\footnotesize{symm}}} \nonumber \\
\left. \left\langle \bar{\psi}(p) \psi(q) A^c_\sigma(r) A^d_\rho(-p-q-r)
\right\rangle \right|_{\mbox{\footnotesize{symm}}} &=&
\left. \Sigma^{cd}_{A\psi\,\sigma \rho}(p,q,r)
\right|_{\mbox{\footnotesize{symm}}} \nonumber \\
\left. \left\langle \bar{c}^a(p) c^b(q) \bar{c}^c(r) c^d(-p-q-r)
\right\rangle \right|_{\mbox{\footnotesize{symm}}} &=&
\left. \Sigma^{abcd}_{cc}(p,q,r) \right|_{\mbox{\footnotesize{symm}}} 
\nonumber \\
\left. \left\langle \bar{c}^a(p) c^b(q) \bar{\psi}(r) \psi(-p-q-r)
\right\rangle \right|_{\mbox{\footnotesize{symm}}} &=&
\left. \Sigma^{ab}_{c\psi}(p,q,r) \right|_{\mbox{\footnotesize{symm}}} 
\nonumber \\
\left. \left\langle \bar{\psi}^{j\beta}_J(p) \psi^{i}_{I\alpha} (q) 
\bar{\psi}^{k\gamma}_K(r) \psi^{l}_{L\delta} (-p-q-r)
\right\rangle \right|_{\mbox{\footnotesize{symm}}} &=& \left. 
\Sigma_{\psi\psi} (p,q,r) \right|_{\mbox{\footnotesize{symm}}}
\label{five4pt}
\end{eqnarray}
and 
\begin{equation}
\left. \left\langle A^a_\mu(p) A^b_\nu(q) A^c_\sigma(r) A^d_\rho(-p-q-r)
\right\rangle \right|_{\mbox{\footnotesize{symm}}} ~=~
\left. \Sigma^{abcd}_{AA\, \mu \nu \sigma \rho}(p,q,r)
\right|_{\mbox{\footnotesize{symm}}} ~.
\end{equation}
We include the quartic gluon one for completeness and for reference to the 
parallel computation in \cite{9} but have appended the label $AA$ to the 
definition used in \cite{9}. The restriction denoted by $\mbox{symm}$ indicates
evaluation with the values of (\ref{symmpt}). In each Green's function we have 
included various indices. These are the adjoint colour indices
$1$~$\leq$~$a$~$\leq$~$(\Nc^2-1)$, colour spinor indices 
$1$~$\leq$~$I$~$\leq$~$\Nc$, Lorentz indices $\mu$, spinor indices 
$1$~$\leq$~$\alpha$~$\leq$~$4$, and flavour indices $1$~$\leq$~$i$~$\leq$~$\Nf$
where $\Nf$ is the number of massless quarks and our colour group generators 
are $T^a_{IJ}$. The colour and Lorentz indices are included in the shorthand 
notation for each Green's function and cannot be treated disjointly. To ease 
discussion we will denote the general form of these by
$\left. \Sigma^{a_1 \ldots a_{n_c}}_{{\cal L}\, \mu_1 \ldots \mu_{n_l}}(p,q,r)
\right|_{\mbox{\footnotesize{symm}}}$ where ${\cal L}$ is the label defining
the Green's function. The convention for the label ${\cal L}$ is that it 
involves a pair of fields. Each field of the pair represents two of the fields 
in the actual $4$-point function in a compact and obvious fashion. We have not 
included the various spinor and flavour indices in the shorthand for 
$\left. \Sigma_{\psi\psi} (p,q,r) \right|_{\mbox{\footnotesize{symm}}}$ as
aside from being cumbersome the computation of this Green's function was 
carried out in a different way to the other five and will be treated 
differently in the discussion. In defining the ranges of the colour group 
indices we have assumed at the outset that we are working with the colour group
$SU(\Nc)$. As discussed in \cite{8,9} this eases the computation of the 
underlying Feynman graphs and in particular the treatment of box graphs. We 
will use the colour group algorithm given in \cite{9} for the evaluation of 
$\left. \Sigma^{abcd}_{Ac\,\sigma \rho}(p,q,r)
\right|_{\mbox{\footnotesize{symm}}}$ and refer the interested reader to that
for more details. However, it will turn out that the final form of this Green's
function will involve various rank $4$ colour tensors which are $f_4^{abcd}$,
$f_4^{adbc}$ and $d_A^{abcd}$ where
\begin{equation}
f_4^{abcd} ~=~ f^{abe} f^{cde} ~~~~,~~~~ 
d_A^{abcd} ~=~ \frac{1}{6} \mbox{Tr} \left( T_A^a T_A^{(b} T_A^c T_A^{d)}
\right)
\end{equation} 
and $f^{abc}$ are the $SU(\Nc)$ structure constants. The tensor $d_A^{abcd}$ is
totally symmetric in its indices and was introduced in \cite{19}. The subscript
$A$ on the group generators indicates that they are in the adjoint
representation. These are the three combinations of tensors which arose in the 
gluon $4$-point function for the graphs involving fields in the adjoint 
representation, \cite{9}. We exclude $f_4^{acdb}$ as it is not independent due 
to the Jacobi identity. In \cite{9} other rank $4$ tensors arose in the quartic
gluon $4$-point function due to closed loops of quarks. As there is no direct
coupling of the quark to the ghost these tensors do not occur in the quartic
ghost or gluon-ghost $4$-point functions at one loop. For the remaining Green's
function we use the purely $SU(\Nc)$ relation for the group generators  
\begin{equation}
T^a_{IJ} T^a_{KL} ~=~ \frac{1}{2} \left[ \delta_{IL} \delta_{KJ} ~-~
\frac{1}{\Nc} \delta_{IJ} \delta_{KL} \right] 
\label{Tdecomp}
\end{equation}
which was useful for graphs involving quarks in the remaining Green's 
functions. Ordinarily the evaluation of $2$- and $3$-point functions in
perturbative QCD is carried out for a general colour group without resorting to
relations such as (\ref{Tdecomp}) which are group specific. This means that in
those cases expressions for QED can be deduced by taking the abelian limit 
where, for instance, the structure constants are formally set to zero. As we 
will be using (\ref{Tdecomp}) for the $4$-point functions of (\ref{five4pt}) 
the QED limit of the resulting expressions cannot be found by this process.
Instead to obtain these a direct evaluation has to be carried out. Given the 
formalism we will use this is not a difficult task and we will provide the one 
loop structure of the Green's functions  
\begin{eqnarray}
\left. \left\langle A_\mu(p) A_\nu(q) A_\sigma(r) A_\rho(-p-q-r)
\right\rangle \right|_{\mbox{\footnotesize{symm}}} &=&
\left. \Sigma^{\QEDs}_{AA\, \mu \nu \sigma \rho}(p,q,r)
\right|_{\mbox{\footnotesize{symm}}} \nonumber \\
\left. \left\langle \bar{\psi}(p) \psi(q) A_\sigma(r) A_\rho(-p-q-r)
\right\rangle \right|_{\mbox{\footnotesize{symm}}} &=&
\left. \Sigma^{\QEDs}_{A\psi\,\sigma \rho}(p,q,r)
\right|_{\mbox{\footnotesize{symm}}} \nonumber \\
\left. \left\langle \bar{\psi}^{j\beta}(p) \psi^{i}_{\alpha} (q) 
\bar{\psi}^{k\gamma}(r) \psi^{l}_{\delta} (-p-q-r)
\right\rangle \right|_{\mbox{\footnotesize{symm}}} &=& \left. 
\Sigma^{\QEDs}_{\psi\psi} (p,q,r) \right|_{\mbox{\footnotesize{symm}}}
\label{qed4pt}
\end{eqnarray} 
where fields without colour group labels are either the photon or electron of
QED. A superscript label is included to avoid confusion of the results with
those of QCD. Included in this set of Green's functions is the photon $4$-point
function and there are no ghost interactions in a linear covariant gauge fixed
QED Lagrangian which we use here.

For the moment we will concentrate our discussion on the evaluation of each of 
the QCD Green's functions. Where necessary we will append comments on 
differences with the QED evaluation. In the case of QCD the number of different
independent colour group tensors which can appear is rather small and is 
determined from the rules noted above. However, what is more involved is the 
treatment of the Lorentz structure. In \cite{9} a projection method was used to
decompose the gluon $4$-point function into the $138$ different possible 
Lorentz structures which could be built from the independent external momenta 
$p$, $q$ and $r$ as well as the metric $\eta^{\mu\nu}$. One of the reasons a 
projection method was necessary rested in the way the individual Feynman graphs
of a Green's function were evaluated. Use was made of the Laporta algorithm 
\cite{20}. This is an integration by parts routine where linear relations are 
derived between scalar Feynman integrals in $d$-dimensions. These can be solved
algebraically in such a way that all the integrals are related to a small set 
of what is termed master integrals. These are determined by direct methods and 
hence the whole evaluation of the Green's function can be coded in a fully 
automatic fashion, \cite{20}. What is key to the application of the Laporta 
algorithm is writing the Green's function of interest in terms of scalar 
Feynman integrals. When the Green's function has Lorentz or spinor indices this
means they first have to be converted to a scalar form. While we do this with a
projection method for most of the Green's functions of interest here, that for 
the quartic quark one is treated differently. This can best be understood by 
first considering the construction of the basis for three of the other four 
Green's functions. By this we mean gluon-ghost, gluon-quark and ghost-quark as 
the quartic ghost $4$-point function does not require a Lorentz decomposition 
as the ghost is a Lorentz singlet. Of these three two have spinor indices but 
not that for the gluon-ghost $4$-point function. For the Lorentz basis of the 
latter the rank $2$ Lorentz tensors of the basis have to be built from the same
basic objects as that for the quartic gluonic one. Rather than the $138$
possibilities for that rank $4$ $4$-point function the basis for the 
gluon-ghost $4$-point function has $10$ independent possibilities. One of these
is $\eta^{\mu\nu}$ while the remaining nine correspond to the $3^2$ independent 
products of the three independent external momenta with different Lorentz
indices. While this is a simple analysis for a quarkless $4$-point function the
basis of objects for a tensor basis has to be increased when there are quark
external legs. This is because one has to allow for spinor indices which are
carried by the unit spinor matrix and the $\gamma$-matrices. In the case of the
latter the $\gamma$-matrices can have free Lorentz indices or be contracted
with external momenta. Additionally there is the complication of amending the
decomposition to allow for the use of dimensional regularization which we use
throughout. In dimensionally regularizing in $d$~$=$~$4$~$-$~$2\epsilon$
dimensions, where $\epsilon$ is the regularizing parameter, the 
$\gamma$-algebra ceases to be finite dimensional and instead is infinite
dimensional. To accommodate this both within the actual computation and the
construction of the tensor basis we use the generalized $\gamma$-matrices
denoted by $\Gamma_{(n)}^{\mu_1 \ldots \mu_n}$, \cite{21,22,23,24,25} , and
defined as the antisymmetric product of the $\gamma$-matrices satisfying the
$d$-dimensional Clifford algebra
\begin{equation}
\{ \gamma^\mu, \gamma^\nu \} ~=~ 2 \eta^{\mu\nu}
\end{equation}
as 
\begin{equation}
\Gamma_{(n)}^{\mu_1 \ldots \mu_n} ~=~ \gamma^{[\mu_1} \ldots \gamma^{\mu_n]} ~.
\end{equation}
A factor of $1/n!$ is understood in the antisymmetrization. We will regard
$\gamma^\mu$ and $\Gamma_{(1)}^\mu$ as being synonymous throughout. A major 
benefit of this choice for the $d$-dimensional $\gamma$-matrices is that the 
$\Gamma_{(n)}$-matrices naturally partition the infinite dimensional spinor
space since the generalized trace is  
\begin{equation}
\mbox{tr} \left( \Gamma_{(m)}^{\mu_1 \ldots \mu_m}
\Gamma_{(n)}^{\nu_1 \ldots \nu_n} \right) ~ \propto ~ \delta_{mn}
I^{\mu_1 \ldots \mu_m \nu_1 \ldots \nu_n} ~.
\end{equation}
The quantity $I^{\mu_1 \ldots \mu_m \nu_1 \ldots \nu_n}$ represents the unit 
matrix in the $\Gamma_{(n)}$-space where there is no sum on $m$ or $n$. As an
aside on the use of dimensional regularization we use the scale $\mu$ of 
(\ref{symmpt}) to ensure that the coupling constant in the regularized theory
is dimensionless. This is the same convention as was used in \cite{9}.  

More importantly this partitioning gives a natural way of constructing a 
Lorentz basis for the gluon-quark and ghost-quark Green's functions. First one
determines which partitions of the $\Gamma_{(n)}$-basis will be present. In
the case of these two Green's functions $n$ will be odd as we are using 
massless quarks and the spinor line will be the product of $\gamma$-matrices
from the quark-gluon vertex Feynman rule or the quark propagator. As there are 
no Lorentz indices on the ghost-quark $4$-point function then the Lorentz 
indices of the contributing $\Gamma_{(n)}$-matrices have to be contracted with 
the independent external momenta. By this reasoning there are only four 
possibilities with $n$ being no larger than $3$. Three derive from the 
contraction of $\Gamma_{(1)}^\mu$ with each external momenta and the final one 
involves $\Gamma_{(3)}^{\mu\nu\sigma}$ contracted with the three different 
momenta due to the antisymmetric property. For the gluon-quark $4$-point 
function the presence of two Lorentz indices increases the number of possible 
tensors in comparison. Also this means that for the $\Gamma_{(n)}$-matrix 
partition $n$ can be at most five since $\Gamma_{(5)}^{\mu\nu pqr}$ can be 
present. For the lower partitions we constructed the basis by considering the 
possible cases for $n$~$=$~$1$ and $3$ separately. When a Lorentz index on the 
$\Gamma_{(n)}$-matrix is contracted with an external momentum the free Lorentz
indices can be carried by the external momenta themselves and the metric
tensor. For the $n$~$=$~$1$ partition there are $36$ tensors but only $31$ for
$n$~$=$~$3$ and $1$ for $n$~$=$~$5$. The full set is recorded in the Appendix
using the notation we will elaborate on shortly. Equipped with this knowledge
of how the tensors for the Lorentz basis are constructed it is easy to see that
its application to the quartic quark Green's function is not straightforward.
For that $4$-point function one has two independent spinor strings and the
spinor sector of the Lorentz basis will formally be
$\Gamma_{(m)}$~$\otimes$~$\Gamma_{(n)}$ where $m$ and $n$ are not necessarily
equal and the tensor product reflects the two independent spinor threads. The 
imbalance is possible due to contractions with independent external momenta as 
will become evident. To construct the basis for this Green's function in 
generality is not practical. For instance, at high enough loop order one would 
have to allow in principle for possible tensors in the basis such as
$\Gamma_{(7)}^{\mu\nu\sigma\rho pqr}$~$\otimes$~$\Gamma_{(5)}^{\mu\nu\sigma\rho
q}$ for example and similar higher partitions. This is virtually impossible to
accommodate in the projection method. Therefore for the quartic quark
$4$-point function we have followed the variation presented in \cite{26}. In 
\cite{26} and \cite{27} a similar problem arose but with three spinor strings. 
To effect the Laporta algorithm scalar integrals were produced by stripping all
the $\gamma$-algebra from the individual Feynman integrals and then decomposing
the underlying Lorentz integrals into scalar integrals coupled with Lorentz 
tensors involving only the metric and external momenta. We refer the interested
reader to \cite{26} for an example of the more technical aspects of this point
which is straightforward to implement at one loop. Beyond that order it becomes
less efficient than the projection method. However, after the Lorentz tensor 
integrals have been decomposed the $\gamma$-matrices are recontracted and then 
these $\gamma$-matrix strings are decomposed into the $\Gamma_{(n)}$-matrix 
basis which is achieved via the recursive relations, \cite{23,24,25}, 
\begin{eqnarray}
\Gamma^{\mu_1 \ldots \mu_n}_{(n)} \gamma^\nu &=&
\Gamma^{\mu_1 \ldots \mu_n \nu}_{(n+1)} ~+~ \sum_{r=1}^n (-1)^{n-r} \,
\eta^{\mu_r \nu} \, \Gamma^{\mu_1 \ldots \mu_{r-1} \mu_{r+1} \ldots
\mu_n}_{(n-1)} \\
\gamma^\nu \Gamma^{\mu_1 \ldots \mu_n}_{(n)} &=&
\Gamma^{\nu \mu_1 \ldots \mu_n}_{(n+1)} ~+~ \sum_{r=1}^n (-1)^{r-1} \,
\eta^{\mu_r \nu} \, \Gamma^{\mu_1 \ldots \mu_{r-1} \mu_{r+1} \ldots
\mu_n}_{(n-1)} ~.
\end{eqnarray}  
The outcome of this detour for the quartic quark Green's function is to arrive
at the point where that $4$-point function is written in terms of scalar
Feynman integrals. 

For the other three Lorentz $4$-point functions the actual projection matrices
which act on each Green's function to produce the analogous scalar Feynman
integrals needs to be constructed. We can treat each of the gluon-ghost, 
gluon-quark and ghost-quark $4$-point functions simultaneously if we write the
decomposition as 
\begin{equation}
\left. \Sigma^{a_1\ldots a_{n_c} \, \mu_1 \ldots \mu_{n_l}}_{{\cal L}\,}(p,q,r)
\right|_{\mbox{\footnotesize{symm}}} ~=~ \sum_{k=1}^{n_p}
{\cal P}_{{\cal L}\, (k)}^{\mu_1 \ldots \mu_{n_l}}(p,q,r) \,
\left. \Sigma^{a_1 \ldots a_{n_c}}_{{\cal L}\,(k)}(p,q,r) 
\right|_{\mbox{\footnotesize{symm}}}
\end{equation}
where ${\cal P}_{{\cal L}\, (k)}^{\mu_1 \ldots \mu_{n_l}}(p,q,r)$ are the
Lorentz tensors of the respective bases and 
$\Sigma^{a_1 \ldots a_{n_c}}_{{\cal L}\,(k)}(p,q,r)$ are the scalar amplitudes
associated with the label $k$ denoting an element of the tensor basis. Each set
is given in the Appendix. The indexing set of the projectors has dimension 
$n_p$ and the number of Lorentz indices is $n_l$. We retain the colour group 
indices on these amplitudes for ease at the moment but the explicit colour 
dependent object will be factored off in the final expression of the Green's 
function. It is the Lorentz tensors which are provided in the Appendix and 
which we use subsequently. To construct the projection matrix we first find the
matrix ${\cal N}_{{\cal L}\, kk^\prime}$ defined by
\begin{equation}
{\cal N}_{{\cal L}\, kk^\prime} ~=~ 
\left. {\cal P}_{{\cal L}\, (k) \, \mu_1 \ldots \mu_{n_l}}(p,q,r) 
{\cal P}^{\mu_1 \ldots \mu_{n_l}}_{{\cal L}\, (k^\prime)}(p,q,r)
\right|_{\mbox{\footnotesize{symm}}} 
\end{equation}
which is symmetric in the Lorentz tensor basis labels $k$ and $k^\prime$. Its 
elements are polynomials in $d$ and its inverse is 
${\cal M}_{{\cal L}\, kk^\prime}$ which is the projector on the Green's 
function. In other words the amplitudes in the Lorentz decomposition are given 
by 
\begin{equation}
\left. \Sigma^{abcd}_{AA\, (k)}(p,q,r) \right|_{\mbox{\footnotesize{symm}}} ~=~
{\cal M}_{AA\, kk^\prime} 
\left( {\cal P}^{\mu \nu \sigma \rho}_{AA\, (k^\prime)}(p,q,r) 
\left.  \left\langle A^a_\mu(p) A^b_\nu(q) A^c_\sigma(r) A^d_\rho(-p-q-r)
\right\rangle \right ) \right|_{\mbox{\footnotesize{symm}}}
\end{equation}
for the quartic gluon Green's function where there is a sum over the label 
$k^\prime$. A summary of the data for the parameters is given in Table $1$ 
where $n_s$ is the number of independent spinor threads for each Green's 
function. We have included the same data for the quartic gluon case for 
completeness as well as for reference to \cite{9}. The parallel data for the
QED analysis is also included for comparison. No information appears in the
$n_c$ column in that case.

{\begin{table}[ht]
\begin{center}
\begin{tabular}{|c||c|c|c|c|c|}
\hline
Green's function & $n_p$ & $n_c$ & $n_l$ & $n_\gamma$ & Graphs \\
\hline
$A^a_\mu A^b_\nu A^c_\sigma A^d_\rho$ & $138$ & $4$ & $4$ & $0$ & $24$ \\ 
$\bar{c}^a c^b A^c_\sigma A^d_\rho$ & $~10$ & $4$ & $2$ & $0$ & $~7$ \\
$\bar{\psi} \psi A^c_\sigma A^d_\rho$ & $~68$ & $2$ & $2$ & $1$ & $~7$ \\
$\bar{c}^a c^b \bar{c}^c c^d$ & $~~\,1$ & $4$ & $0$ & $0$ & $~4$ \\
$\bar{c}^a c^b \bar{\psi} \psi$ & $~~\,4$ & $2$ & $0$ & $1$ & $~2$ \\
$\bar{\psi}^{j\beta}_J \psi^{i}_{I\alpha} \bar{\psi}^{k\gamma}_K 
\psi^{l}_{L\delta}$ & $(168)$ & $0$ & $0$ & $2$ & $~4$ \\
\hline
$A_\mu A_\nu A_\sigma A_\rho$ & $138$ & $-$ & $4$ & $0$ & $~6$ \\ 
$\bar{\psi} \psi A_\sigma A_\rho$ & $~68$ & $-$ & $2$ & $1$ & $~2$ \\
$\bar{\psi}^{j\beta} \psi^{i}_{\alpha} \bar{\psi}^{k\gamma} \psi^{l}_{\delta}$ &
$~(72)$ & $-$ & $0$ & $2$ & $~4$ \\
\hline
\end{tabular}
\end{center}
\begin{center}
{Table 1. Summary data for each of the Green's functions in QCD (top) and QED
(bottom).}
\end{center}
\end{table}}

The final column in Table $1$ indicates the number of one loop Feynman graphs
contributing to each Green's function. The absence of a ghost in a QED linear 
covariant gauge fixing reduces the number of Feynman graphs needed to be 
computed directly in that case. This leads us to briefly discuss the technical 
machinery used to evaluate all the $4$-point functions to one loop. The Feynman
graphs are generated automatically by the {\sc Qgraf} package, \cite{28}, in an
electronic format which can be adapted to the symbolic manipulation language 
{\sc Form}, \cite{29,30}, which we have used to write the above algorithm in. 
We have encoded both the projection method and the method to write the quark 
$4$-point function as scalar Feynman integrals in separate {\sc Form} modules 
after all the Lorentz, spinor and colour group indices have been mapped into 
the {\sc Qgraf} generated graphs. In each of the approaches scalar products of 
the internal momentum and the external momenta are rewritten in terms of the 
propagators of the integral. As this is a one loop computation for a $4$-point 
function there are no irreducible scalar products although there are three 
different box integral families at one loop. By integral family we mean the 
core topologies, with the constraints (\ref{symmpt}) implemented, in the syntax
of the Laporta algorithm, \cite{20}. To effect the Laporta algorithm we used 
its implementation in the early version of the package {\sc Reduze}, \cite{31},
which was also used in \cite{9}. The package has the feature that the relations
giving the reduction to the master integrals can be converted to {\sc Form} 
notation. So we have written these in a {\sc Form} module and included it 
within our automatic Feynman graph computation. Once all the projections and 
integration by parts for a graph have been applied the final stage is the 
substitution of the various master integrals. For the symmetric point, 
(\ref{symmpt}), their explicit values were recorded in \cite{9}. However, these
were special cases of the more general evaluation of one loop massless triangle
and box integrals derived in \cite{32,33,34,35} for all external legs fully 
off-shell. 

This completes the technical aspects of the tools we used to implement the 
algorithm to evaluate the Green's functions.  One minor feature of our 
computations which differs from the quartic gluon Green's function is that that
$4$-point function needed to be renormalized. This is not the case here for the
QED and the QCD $4$-point functions of (\ref{qed4pt}) and (\ref{five4pt}) as 
they have no corresponding interaction in the original QED or QCD Lagrangians 
with a linear covariant gauge fixing. So they all have to be finite without the
introduction of any new renormalization constant. This provides a minor check 
on our computation and we indeed found that all the one loop Green's functions 
were finite. At higher loop the wave function renormalization for the external 
legs of the Green's functions and the conversion of the bare parameters to 
renormalized variables would have to be included. At one loop this is not 
necessary. It is also worth noting that these comments have to be modified if 
one uses a different gauge fixing to the canonical linear gauge fixing. For 
instance, in the nonlinear Curci-Ferrari gauge, \cite{18}, there is a quartic 
ghost interaction in the gauge fixed Lagrangian. Equally in the maximal abelian
gauge, \cite{36,37,38}, there are gluon-ghost $4$-point interactions in 
addition to quartic ghost ones. Of course in both these gauges the final form 
of the various Green's functions we consider here would be structurally 
different.

\sect{QCD results.}

We are now in a position to discuss the results of the computations. As there
are no Lorentz structures in the ghost $4$-point function we can illustrate
main aspects of the form of the results for this and the other Green's 
function. For an arbitrary linear covariant gauge we find 
\begin{eqnarray}
\left. \Sigma^{abcd}_{cc}(p,q,r) \right|_{\mbox{\footnotesize{symm}}} &=&
\left[ \frac{31}{320}
- \frac{51}{160} \alpha
+ \frac{71}{320} \alpha^2
+ \frac{177}{1600} \ln\left(\frac{4}{3}\right)
- \frac{107}{800} \ln\left(\frac{4}{3}\right) \alpha
\right. \nonumber \\
&& \left. ~
+ \frac{37}{1600} \ln\left(\frac{4}{3}\right) \alpha^2
- \frac{1003}{51200} \Phi_1\left(\frac{9}{16},\frac{9}{16}\right)
+ \frac{423}{25600} \Phi_1\left(\frac{9}{16},\frac{9}{16}\right) \alpha
\right. \nonumber \\
&& \left. ~
+ \frac{3357}{51200} \Phi_1\left(\frac{9}{16},\frac{9}{16}\right) \alpha^2
- \frac{1}{32} \Phi_1\left(\frac{3}{4},\frac{3}{4}\right)
+ \frac{5}{64} \Phi_1\left(\frac{3}{4},\frac{3}{4}\right) \alpha
\right. \nonumber \\
&& \left. ~
- \frac{11}{64} \Phi_1\left(\frac{3}{4},\frac{3}{4}\right) \alpha^2
\right]
f_4^{acbd} \Nc a ~+~ O(a^2) 
\end{eqnarray}
where $a$~$=$~$g^2/(16\pi^2)$ is the renormalized coupling constant and 
$\alpha$ is the renormalized gauge fixing parameter of the linear covariant 
gauge fixing we use throughout. The Landau gauge corresponds to 
$\alpha$~$=$~$0$. The function $\Phi_1(x,y)$ corresponds with $\Phi(x,y)$ of 
\cite{35} and involves the dilogarithm function $\mbox{Li}_2(z)$. The 
definition is
\begin{equation}
\Phi_1(x,y) ~=~ \frac{1}{\lambda} \left[ 2 \mbox{Li}_2(-\rho x)
+ 2 \mbox{Li}_2(-\rho y)
+ \ln \left( \frac{y}{x} \right)
\ln \left( \frac{(1+\rho y)}{(1+\rho x)} \right)
+ \ln(\rho x) \ln(\rho y) + \frac{\pi^2}{3} \right]
\end{equation}
where 
\begin{equation}
\lambda(x,y) ~=~ \sqrt{\Delta_G} ~~~,~~~
\rho(x,y) ~=~ \frac{2}{[1-x-y+\lambda(x,y)]}
\end{equation}
and
\begin{equation}
\Delta_G(x,y) ~=~ x^2 ~-~ 2 x y ~+~ y^2 ~-~ 2 x ~-~ 2 y ~+~ 1
\end{equation}
which is the Gram determinant. The function $\Phi_1(x,y)$ appears throughout 
all the Green's functions and arises in two contexts in each computation. They
occur in masters which are in effect a triangle or $3$-point function and when 
there is a pure box graph. The triangle master emerges either as a consequence
of the original Feynman graph being a triangle before integration by parts
reduction or within the reduction of an original box graph. Each function is 
reflected differently in the analytic form of the Green's function at the 
symmetric point as $\Phi_1 \left( \frac{3}{4}, \frac{3}{4} \right)$ and  
$\Phi_1 \left( \frac{9}{16}, \frac{9}{16} \right)$. The reason for the 
different arguments is due to the symmetric point values of the squared 
external momenta, \cite{32,33,34,35}. More specifically the two functions are
related to the Clausen function ${\mbox{Cl}}(\theta)$ by, \cite{32,33}, 
\begin{eqnarray}
\Phi_1 \left( \frac{3}{4},\frac{3}{4} \right) &=& \sqrt{2} \left[
2 \mbox{Cl}_2 \left( 2 \cos^{-1} \left( \frac{1}{\sqrt{3}} \right) \right)
+ \mbox{Cl}_2 \left( 2 \cos^{-1} \left( \frac{1}{3} \right) \right) \right]
\nonumber \\
\Phi_1 \left( \frac{9}{16},\frac{9}{16} \right) &=& \frac{4}{\sqrt{5}} \left[
2 \mbox{Cl}_2 \left( 2 \cos^{-1} \left( \frac{2}{3} \right) \right)
+ \mbox{Cl}_2 \left( 2 \cos^{-1} \left( \frac{1}{9} \right) \right) \right] ~.
\end{eqnarray}
While we have given the quartic ghost function explicitly we give a flavour of
the remaining functions by recording their numerical values. We note that the 
full analytic form of each of the $4$-point functions given here are provided 
in an attached data file. For the numerical versions we note that  
\begin{equation}
\Phi_1 \left( \frac{3}{4}, \frac{3}{4} \right) ~=~ 2.832045 ~~~,~~~
\Phi_1 \left( \frac{9}{16}, \frac{9}{16} \right) ~=~ 3.403614 ~.
\end{equation}

With these values it is straightforward to provide the Green's functions in a 
compact form. All the cases we record explicitly are for the $SU(3)$ colour 
group. First, for comparison the quartic ghost function is
\begin{equation}
\left. \Sigma^{abcd}_{cc}(p,q,r) \right|_{\mbox{\footnotesize{symm}}} ~=~ -~
\left[ 0.079434 + 0.239204 \alpha + 0.105202 \alpha^2 \right]
f_4^{acbd} a ~+~ O(a^2)
\end{equation}
in an arbitrary gauge. As the remaining $4$-point functions are much larger the
remaining Green's functions are all given in the Landau gauge. For the 
gluon-ghost $4$-point function we have 
\begin{eqnarray}
\left. \Sigma^{abcd}_{Ac\,\sigma \rho}(p,q,r)
\right|_{\mbox{\footnotesize{symm}}} &=& 
\left[ 
  0.069633 {\cal P}_{Ac\, (1)}
- 0.018972 {\cal P}_{Ac\, (2)}
- 0.566599 {\cal P}_{Ac\, (3)}
\right. \nonumber \\
&& \left. ~
+ 0.091060 {\cal P}_{Ac\, (4)}
- 0.445930 {\cal P}_{Ac\, (5)}
+ 0.101697 {\cal P}_{Ac\, (6)}
\right. \nonumber \\
&& \left. ~
+ 0.211729 {\cal P}_{Ac\, (7)}
- 0.048584 {\cal P}_{Ac\, (8)}
- 0.169253 {\cal P}_{Ac\, (9)}
\right. \nonumber \\
&& \left. ~
+ 0.042475 {\cal P}_{Ac\, (10)}
\right] d_A^{abcd} a 
\nonumber \\
&& +~ \left[ 
 0.192709 {\cal P}_{Ac\, (1)}
+ 1.000682 {\cal P}_{Ac\, (2)}
- 0.506828 {\cal P}_{Ac\, (3)}
\right. \nonumber \\
&& \left. ~~~~
+ 1.115084 {\cal P}_{Ac\, (4)}
- 0.632469 {\cal P}_{Ac\, (5)}
- 0.811897 {\cal P}_{Ac\, (6)}
\right. \nonumber \\
&& \left. ~~~~
- 0.697495 {\cal P}_{Ac\, (7)}
+ 0.595902 {\cal P}_{Ac\, (8)}
- 0.853676 {\cal P}_{Ac\, (9)}
\right. \nonumber \\
&& \left. ~~~~
+ 0.024346 {\cal P}_{Ac\, (10)}
\right] f_4^{abcd} a 
\nonumber \\
&& +~ \left[ 
  0.385419 {\cal P}_{Ac\, (1)}
+ 0.314724 {\cal P}_{Ac\, (2)}
- 1.013359 {\cal P}_{Ac\, (3)}
\right. \nonumber \\
&& \left. ~~~~
+ 0.543528 {\cal P}_{Ac\, (4)}
- 1.376360 {\cal P}_{Ac\, (5)}
- 0.048277 {\cal P}_{Ac\, (6)}
\right. \nonumber \\
&& \left. ~~~~
+ 0.180527 {\cal P}_{Ac\, (7)}
- 0.494836 {\cal P}_{Ac\, (8)}
- 0.131835 {\cal P}_{Ac\, (9)}
\right. \nonumber \\
&& \left. ~~~~
+ 0.048692 {\cal P}_{Ac\, (10)}
\right] f_4^{adbc} a ~+~ O(a^2) ~.
\end{eqnarray}
This form is indicative of the other Green's function in that the colour group 
structure is more involved. We have not implemented the Jacobi identity for 
instance to rearrange the expression as it was not clear if this would lead to 
any simplifications. Moreover, with two different external fields with one 
external momentum expressed in terms of the others the symmetry structure is 
not apparent. For the gluon-quark $4$-point function not every Lorentz 
structure is present in each colour group channel as  
\begin{eqnarray}
\left. \Sigma^{cd}_{A\psi\,\sigma \rho}(p,q,r)
\right|_{\mbox{\footnotesize{symm}}} &=&
\left[ 
 0.051996 {\cal P}_{A\psi\, (1)}
- 0.051996 {\cal P}_{A\psi\, (2)}
+ 0.051996 {\cal P}_{A\psi\, (4)}
\right. \nonumber \\
&& \left. ~
- 0.051996 {\cal P}_{A\psi\, (5)}
+ 0.665903 {\cal P}_{A\psi\, (7)}
- 0.665903 {\cal P}_{A\psi\, (8)}
\right. \nonumber \\
&& \left. ~
+ 0.052679 {\cal P}_{A\psi\, (10)}
+ 0.625933 {\cal P}_{A\psi\, (11)}
+ 0.728409 {\cal P}_{A\psi\, (12)}
\right. \nonumber \\
&& \left. ~
- 0.516160 {\cal P}_{A\psi\, (13)}
+ 0.054818 {\cal P}_{A\psi\, (14)}
+ 1.821741 {\cal P}_{A\psi\, (15)}
\right. \nonumber \\
&& \left. ~
- 0.262442 {\cal P}_{A\psi\, (16)}
+ 0.241748 {\cal P}_{A\psi\, (17)}
+ 1.264728 {\cal P}_{A\psi\, (18)}
\right. \nonumber \\
&& \left. ~
- 0.054818 {\cal P}_{A\psi\, (19)}
+ 0.516160 {\cal P}_{A\psi\, (20)}
- 1.821741 {\cal P}_{A\psi\, (21)}
\right. \nonumber \\
&& \left. ~
- 0.625933 {\cal P}_{A\psi\, (22)}
- 0.052679 {\cal P}_{A\psi\, (23)}
- 0.728409 {\cal P}_{A\psi\, (24)}
\right. \nonumber \\
&& \left. ~
- 0.241748 {\cal P}_{A\psi\, (25)}
+ 0.262442 {\cal P}_{A\psi\, (26)}
- 1.264728 {\cal P}_{A\psi\, (27)}
\right. \nonumber \\
&& \left. ~
- 0.798761 {\cal P}_{A\psi\, (28)}
+ 1.436663 {\cal P}_{A\psi\, (29)}
- 0.798761 {\cal P}_{A\psi\, (30)}
\right. \nonumber \\
&& \left. ~
- 1.436663 {\cal P}_{A\psi\, (31)}
+ 0.798761 {\cal P}_{A\psi\, (32)}
+ 0.798761 {\cal P}_{A\psi\, (33)}
\right. \nonumber \\
&& \left. ~
- 0.798761 {\cal P}_{A\psi\, (34)}
+ 0.798761 {\cal P}_{A\psi\, (35)}
- 1.302123 {\cal P}_{A\psi\, (37)}
\right. \nonumber \\
&& \left. ~
- 1.302123 {\cal P}_{A\psi\, (38)}
- 2.604247 {\cal P}_{A\psi\, (39)}
- 0.376857 {\cal P}_{A\psi\, (40)}
\right. \nonumber \\
&& \left. ~
+ 0.376857 {\cal P}_{A\psi\, (41)}
- 0.764445 {\cal P}_{A\psi\, (43)}
- 1.272049 {\cal P}_{A\psi\, (44)}
\right. \nonumber \\
&& \left. ~
- 1.016975 {\cal P}_{A\psi\, (45)}
+ 1.272049 {\cal P}_{A\psi\, (46)}
+ 0.764445 {\cal P}_{A\psi\, (47)}
\right. \nonumber \\
&& \left. ~
+ 1.016975 {\cal P}_{A\psi\, (48)}
+ 0.130747 {\cal P}_{A\psi\, (49)}
- 0.130747 {\cal P}_{A\psi\, (50)}
\right. \nonumber \\
&& \left. ~
- 0.252531 {\cal P}_{A\psi\, (52)}
+ 0.255073 {\cal P}_{A\psi\, (53)}
- 1.016975 {\cal P}_{A\psi\, (54)}
\right. \nonumber \\
&& \left. ~
- 0.255073 {\cal P}_{A\psi\, (55)}
+ 0.252531 {\cal P}_{A\psi\, (56)}
+ 1.016975 {\cal P}_{A\psi\, (57)}
\right. \nonumber \\
&& \left. ~
- 0.258102 {\cal P}_{A\psi\, (59)}
- 0.343551 {\cal P}_{A\psi\, (60)}
- 0.258102 {\cal P}_{A\psi\, (61)}
\right. \nonumber \\
&& \left. ~
+ 0.343551 {\cal P}_{A\psi\, (62)}
+ 0.258102 {\cal P}_{A\psi\, (63)}
+ 0.258102 {\cal P}_{A\psi\, (64)}
\right. \nonumber \\
&& \left. ~
- 0.258102 {\cal P}_{A\psi\, (65)}
+ 0.258102 {\cal P}_{A\psi\, (66)}
- 1.701880 {\cal P}_{A\psi\, (68)}
\right] \delta^{cd} a \nonumber \\ 
&& + \left[ 
0.029665 {\cal P}_{A\psi\, (1)}
+ 2.316733 {\cal P}_{A\psi\, (2)}
+ 2.344135 {\cal P}_{A\psi\, (3)}
\right. \nonumber \\
&& \left. ~~~
+ 0.027402 {\cal P}_{A\psi\, (4)}
+ 2.314469 {\cal P}_{A\psi\, (5)}
+ 2.344135 {\cal P}_{A\psi\, (6)}
\right. \nonumber \\
&& \left. ~~~
- 0.066898 {\cal P}_{A\psi\, (7)}
- 3.976814 {\cal P}_{A\psi\, (8)}
- 4.043712 {\cal P}_{A\psi\, (9)}
\right. \nonumber \\
&& \left. ~~~
+ 0.838848 {\cal P}_{A\psi\, (10)}
+ 2.284266 {\cal P}_{A\psi\, (11)}
+ 2.172543 {\cal P}_{A\psi\, (12)}
\right. \nonumber \\
&& \left. ~~~
+ 0.727540 {\cal P}_{A\psi\, (13)}
+ 1.579463 {\cal P}_{A\psi\, (14)}
+ 4.425123 {\cal P}_{A\psi\, (15)}
\right. \nonumber \\
&& \left. ~~~
+ 0.450090 {\cal P}_{A\psi\, (16)}
+ 1.223824 {\cal P}_{A\psi\, (17)}
+ 2.342749 {\cal P}_{A\psi\, (18)}
\right. \nonumber \\
&& \left. ~~~
- 0.912599 {\cal P}_{A\psi\, (19)}
+ 0.410488 {\cal P}_{A\psi\, (20)}
- 3.979770 {\cal P}_{A\psi\, (21)}
\right. \nonumber \\
&& \left. ~~~
- 1.005854 {\cal P}_{A\psi\, (22)}
+ 0.533701 {\cal P}_{A\psi\, (23)}
- 1.021505 {\cal P}_{A\psi\, (24)}
\right. \nonumber \\
&& \left. ~~~
- 0.754487 {\cal P}_{A\psi\, (25)}
+ 0.724932 {\cal P}_{A\psi\, (26)}
- 4.308456 {\cal P}_{A\psi\, (27)}
\right. \nonumber \\
&& \left. ~~~
- 2.639334 {\cal P}_{A\psi\, (28)}
+ 1.641136 {\cal P}_{A\psi\, (29)}
- 2.860845 {\cal P}_{A\psi\, (30)}
\right. \nonumber \\
&& \left. ~~~
- 2.810777 {\cal P}_{A\psi\, (31)}
+ 1.092665 {\cal P}_{A\psi\, (32)}
+ 0.871154 {\cal P}_{A\psi\, (33)}
\right. \nonumber \\
&& \left. ~~~
- 2.836861 {\cal P}_{A\psi\, (34)}
+ 0.895138 {\cal P}_{A\psi\, (35)}
- 1.965707 {\cal P}_{A\psi\, (36)}
\right. \nonumber \\
&& \left. ~~~
- 4.789443 {\cal P}_{A\psi\, (37)}
- 2.415426 {\cal P}_{A\psi\, (38)}
- 7.204870 {\cal P}_{A\psi\, (39)}
\right. \nonumber \\
&& \left. ~~~
+ 0.630272 {\cal P}_{A\psi\, (40)}
+ 1.068829 {\cal P}_{A\psi\, (41)}
- 2.643495 {\cal P}_{A\psi\, (42)}
\right. \nonumber \\
&& \left. ~~~
+ 1.033312 {\cal P}_{A\psi\, (43)}
- 0.384262 {\cal P}_{A\psi\, (44)}
- 0.562489 {\cal P}_{A\psi\, (45)}
\right. \nonumber \\
&& \left. ~~~
+ 0.717403 {\cal P}_{A\psi\, (46)}
- 0.210354 {\cal P}_{A\psi\, (47)}
+ 1.420985 {\cal P}_{A\psi\, (48)}
\right. \nonumber \\
&& \left. ~~~
- 2.259213 {\cal P}_{A\psi\, (49)}
- 4.165986 {\cal P}_{A\psi\, (50)}
- 1.784999 {\cal P}_{A\psi\, (51)}
\right. \nonumber \\
&& \left. ~~~
- 1.631339 {\cal P}_{A\psi\, (52)}
- 0.703582 {\cal P}_{A\psi\, (53)}
- 1.420985 {\cal P}_{A\psi\, (54)}
\right. \nonumber \\
&& \left. ~~~
+ 0.178227 {\cal P}_{A\psi\, (55)}
+ 1.595801 {\cal P}_{A\psi\, (56)}
+ 0.562489 {\cal P}_{A\psi\, (57)}
\right. \nonumber \\
&& \left. ~~~
+ 2.334275 {\cal P}_{A\psi\, (58)}
- 0.490947 {\cal P}_{A\psi\, (59)}
- 0.439694 {\cal P}_{A\psi\, (60)}
\right. \nonumber \\
&& \left. ~~~
- 0.493727 {\cal P}_{A\psi\, (61)}
- 0.273666 {\cal P}_{A\psi\, (62)}
- 0.269298 {\cal P}_{A\psi\, (63)}
\right. \nonumber \\
&& \left. ~~~
- 0.272078 {\cal P}_{A\psi\, (64)}
+ 0.021890 {\cal P}_{A\psi\, (65)}
+ 0.243539 {\cal P}_{A\psi\, (66)}
\right. \nonumber \\
&& \left. ~~~
- 0.250188 {\cal P}_{A\psi\, (67)}
+ 0.319089 {\cal P}_{A\psi\, (68)}
\right] T^c T^d a \nonumber \\ 
&& + \left[ 
-~ 2.316733 {\cal P}_{A\psi\, (1)}
- 0.029665 {\cal P}_{A\psi\, (2)}
- 2.344135 {\cal P}_{A\psi\, (3)}
\right. \nonumber \\
&& \left. ~~~
- 2.314469 {\cal P}_{A\psi\, (4)}
- 0.027402 {\cal P}_{A\psi\, (5)}
- 2.344135 {\cal P}_{A\psi\, (6)}
\right. \nonumber \\
&& \left. ~~~
+ 3.976814 {\cal P}_{A\psi\, (7)}
+ 0.066898 {\cal P}_{A\psi\, (8)}
+ 4.043712 {\cal P}_{A\psi\, (9)}
\right. \nonumber \\
&& \left. ~~~
- 0.533701 {\cal P}_{A\psi\, (10)}
+ 1.005854 {\cal P}_{A\psi\, (11)}
+ 1.021504 {\cal P}_{A\psi\, (12)}
\right. \nonumber \\
&& \left. ~~~
- 0.410488 {\cal P}_{A\psi\, (13)}
+ 0.912599 {\cal P}_{A\psi\, (14)}
+ 3.979770 {\cal P}_{A\psi\, (15)}
\right. \nonumber \\
&& \left. ~~~
- 0.724932 {\cal P}_{A\psi\, (16)}
+ 0.754487 {\cal P}_{A\psi\, (17)}
+ 4.308456 {\cal P}_{A\psi\, (18)}
\right. \nonumber \\
&& \left. ~~~
- 1.579463 {\cal P}_{A\psi\, (19)}
- 0.727540 {\cal P}_{A\psi\, (20)}
- 4.425123 {\cal P}_{A\psi\, (21)}
\right. \nonumber \\
&& \left. ~~~
- 2.284266 {\cal P}_{A\psi\, (22)}
- 0.838848 {\cal P}_{A\psi\, (23)}
- 2.172543 {\cal P}_{A\psi\, (24)}
\right. \nonumber \\
&& \left. ~~~
- 1.223824 {\cal P}_{A\psi\, (25)}
- 0.450090 {\cal P}_{A\psi\, (26)}
- 2.342749 {\cal P}_{A\psi\, (27)}
\right. \nonumber \\
&& \left. ~~~
- 1.092665 {\cal P}_{A\psi\, (28)}
+ 2.810777 {\cal P}_{A\psi\, (29)}
- 0.871154 {\cal P}_{A\psi\, (30)}
\right. \nonumber \\
&& \left. ~~~
- 1.641136 {\cal P}_{A\psi\, (31)}
+ 2.639334 {\cal P}_{A\psi\, (32)}
+ 2.860845 {\cal P}_{A\psi\, (33)}
\right. \nonumber \\
&& \left. ~~~
- 0.895138 {\cal P}_{A\psi\, (34)}
+ 2.836861 {\cal P}_{A\psi\, (35)}
+ 1.965707 {\cal P}_{A\psi\, (36)}
\right. \nonumber \\
&& \left. ~~~
- 2.415426 {\cal P}_{A\psi\, (37)}
- 4.789443 {\cal P}_{A\psi\, (38)}
- 7.204870 {\cal P}_{A\psi\, (39)}
\right. \nonumber \\
&& \left. ~~~
- 1.068829 {\cal P}_{A\psi\, (40)}
- 0.630272 {\cal P}_{A\psi\, (41)}
+ 2.643495 {\cal P}_{A\psi\, (42)}
\right. \nonumber \\
&& \left. ~~~
+ 0.210354 {\cal P}_{A\psi\, (43)}
- 0.717403 {\cal P}_{A\psi\, (44)}
- 1.420985 {\cal P}_{A\psi\, (45)}
\right. \nonumber \\
&& \left. ~~~
+ 0.384262 {\cal P}_{A\psi\, (46)}
- 1.033312 {\cal P}_{A\psi\, (47)}
+ 0.562489 {\cal P}_{A\psi\, (48)}
\right. \nonumber \\
&& \left. ~~~
+ 4.165986 {\cal P}_{A\psi\, (49)}
+ 2.259213 {\cal P}_{A\psi\, (50)}
+ 1.784999 {\cal P}_{A\psi\, (51)}
\right. \nonumber \\
&& \left. ~~~
- 1.595801 {\cal P}_{A\psi\, (52)}
- 0.178227 {\cal P}_{A\psi\, (53)}
- 0.562489 {\cal P}_{A\psi\, (54)}
\right. \nonumber \\
&& \left. ~~~
+ 0.703582 {\cal P}_{A\psi\, (55)}
+ 1.631339 {\cal P}_{A\psi\, (56)}
+ 1.420985 {\cal P}_{A\psi\, (57)}
\right. \nonumber \\
&& \left. ~~~
- 2.334275 {\cal P}_{A\psi\, (58)}
+ 0.269298 {\cal P}_{A\psi\, (59)}
+ 0.273666 {\cal P}_{A\psi\, (60)}
\right. \nonumber \\
&& \left. ~~~
+ 0.272078 {\cal P}_{A\psi\, (61)}
+ 0.439694 {\cal P}_{A\psi\, (62)}
+ 0.490947 {\cal P}_{A\psi\, (63)}
\right. \nonumber \\
&& \left. ~~~
+ 0.493727 {\cal P}_{A\psi\, (64)}
- 0.243539 {\cal P}_{A\psi\, (65)}
- 0.021890 {\cal P}_{A\psi\, (66)}
\right. \nonumber \\
&& \left. ~~~
+ 0.250188 {\cal P}_{A\psi\, (67)}
+ 0.319089 {\cal P}_{A\psi\, (68)}
\right] T^d T^c a ~+~ O(a^2) 
\label{qcdaq}
\end{eqnarray}
where the group generators are present. A degree of symmetry can be observed.
For instance, in the first colour channel quite a few of the Lorentz channels 
can be paired asymmetrically. The last projected $4$-point function is the 
ghost-quark $4$-point function which is 
\begin{eqnarray}
\left. \Sigma^{ab}_{c\psi}(p,q,r) \right|_{\mbox{\footnotesize{symm}}} &=&
\left[ 
 0.112015 {\cal P}_{c\psi\, (1)}
+ 0.112015 {\cal P}_{c\psi\, (2)}
+ 0.224031 {\cal P}_{c\psi\, (3)}
\right] \delta^{ab} a \nonumber \\
&& + \left[ 
0.060233 {\cal P}_{c\psi\, (1)}
+ 0.611859 {\cal P}_{c\psi\, (2)}
+ 0.672092 {\cal P}_{c\psi\, (3)}
\right. \nonumber \\
&& \left. ~~~
+ 0.557425 {\cal P}_{c\psi\, (4)}
\right] T^a T^b a \nonumber \\
&& + \left[ 
0.611859 {\cal P}_{c\psi\, (1)}
+ 0.060233 {\cal P}_{c\psi\, (2)}
+ 0.672092 {\cal P}_{c\psi\, (3)}
\right. \nonumber \\
&& \left. ~~~
- 0.557425 {\cal P}_{c\psi\, (4)}
\right] T^b T^a a ~+~ O(a^2) 
\end{eqnarray}
where the unit matrix in the flavour indices is omitted similar to the previous
Green's function. Also the expression is like (\ref{qcdaq}) as it carries a 
remnant of the underlying symmetry of working at the fully symmetric point. 

As noted the final Green's function was computed in a different way from the 
previous four in that there was no projection on to the Lorentz basis due to 
the presence of two $\gamma$-matrix strings. Instead the one loop computation 
was performed directly and we found 
\begin{eqnarray}
\left. \Sigma_{\psi\psi} (p,q,r) \right|_{\mbox{\footnotesize{symm}}} &=&
\left[ \,
-~ 0.082685 {\cal P}_{\psi\psi\, (1)}
- 0.027899 {\cal P}_{\psi\psi\, (2)}
+ 0.208213 {\cal P}_{\psi\psi\, (3)}
\right. \nonumber \\
&& \left. ~
+~ 0.285516 {\cal P}_{\psi\psi\, (4)}
+ 0.126257 {\cal P}_{\psi\psi\, (5)}
+ 0.255432 {\cal P}_{\psi\psi\, (6)}
\right. \nonumber \\
&& \left. ~
+~ 0.003043 {\cal P}_{\psi\psi\, (7)}
- 0.088114 {\cal P}_{\psi\psi\, (8)}
- 0.114487 {\cal P}_{\psi\psi\, (9)}
\right. \nonumber \\
&& \left. ~
+~ 0.038162 {\cal P}_{\psi\psi\, (10)}
- 0.427903 {\cal P}_{\psi\psi\, (11)}
- 0.252736 {\cal P}_{\psi\psi\, (12)}
\right. \nonumber \\
&& \left. ~
+~ 0.313416 {\cal P}_{\psi\psi\, (13)}
+ 0.290898 {\cal P}_{\psi\psi\, (14)}
- 0.049365 {\cal P}_{\psi\psi\, (15)}
\right. \nonumber \\
&& \left. ~
-~ 0.036201 {\cal P}_{\psi\psi\, (16)}
+ 0.140945 {\cal P}_{\psi\psi\, (17)}
+ 0.164419 {\cal P}_{\psi\psi\, (18)}
\right. \nonumber \\
&& \left. ~
-~ 0.255432 {\cal P}_{\psi\psi\, (19)}
- 0.126257 {\cal P}_{\psi\psi\, (20)}
- 0.126863 {\cal P}_{\psi\psi\, (21)}
\right. \nonumber \\
&& \left. ~
+~ 0.042288 {\cal P}_{\psi\psi\, (22)}
- 0.208942 {\cal P}_{\psi\psi\, (23)}
- 0.283331 {\cal P}_{\psi\psi\, (24)}
\right. \nonumber \\
&& \left. ~
-~ 0.290898 {\cal P}_{\psi\psi\, (25)}
- 0.313416 {\cal P}_{\psi\psi\, (26)}
+ 0.003043 {\cal P}_{\psi\psi\, (27)}
\right. \nonumber \\
&& \left. ~
-~ 0.088114 {\cal P}_{\psi\psi\, (28)}
+ 0.043794 {\cal P}_{\psi\psi\, (29)}
+ 0.144571 {\cal P}_{\psi\psi\, (30)}
\right. \nonumber \\
&& \left. ~
-~ 0.164419 {\cal P}_{\psi\psi\, (31)}
- 0.140945 {\cal P}_{\psi\psi\, (32)}
- 0.039244 {\cal P}_{\psi\psi\, (33)}
\right. \nonumber \\
&& \left. ~
+~ 0.038749 {\cal P}_{\psi\psi\, (34)}
- 0.144571 {\cal P}_{\psi\psi\, (35)}
- 0.043794 {\cal P}_{\psi\psi\, (36)}
\right. \nonumber \\
&& \left. ~
+~ 0.283331 {\cal P}_{\psi\psi\, (37)}
+ 0.208942 {\cal P}_{\psi\psi\, (38)}
+ 0.038749 {\cal P}_{\psi\psi\, (39)}
\right. \nonumber \\
&& \left. ~
-~ 0.039244 {\cal P}_{\psi\psi\, (40)}
- 0.285516 {\cal P}_{\psi\psi\, (41)}
- 0.208213 {\cal P}_{\psi\psi\, (42)}
\right. \nonumber \\
&& \left. ~
-~ 0.038749 {\cal P}_{\psi\psi\, (43)}
+ 0.039244 {\cal P}_{\psi\psi\, (44)}
+ 0.252736 {\cal P}_{\psi\psi\, (45)}
\right. \nonumber \\
&& \left. ~
+~ 0.427903 {\cal P}_{\psi\psi\, (46)}
+ 0.039244 {\cal P}_{\psi\psi\, (47)}
- 0.038749 {\cal P}_{\psi\psi\, (48)}
\right. \nonumber \\
&& \left. ~
-~ 0.038162 {\cal P}_{\psi\psi\, (49)}
+ 0.114487 {\cal P}_{\psi\psi\, (50)}
- 0.036201 {\cal P}_{\psi\psi\, (51)}
\right. \nonumber \\
&& \left. ~
-~ 0.049365 {\cal P}_{\psi\psi\, (52)}
+ 0.027899 {\cal P}_{\psi\psi\, (53)}
+ 0.082685 {\cal P}_{\psi\psi\, (54)}
\right. \nonumber \\
&& \left. ~
-~ 0.088114 {\cal P}_{\psi\psi\, (55)}
+ 0.003043 {\cal P}_{\psi\psi\, (56)}
- 0.088114 {\cal P}_{\psi\psi\, (57)}
\right. \nonumber \\
&& \left. ~
+~ 0.003043 {\cal P}_{\psi\psi\, (58)}
+ 0.042288 {\cal P}_{\psi\psi\, (59)}
- 0.126863 {\cal P}_{\psi\psi\, (60)}
\right. \nonumber \\
&& \left. ~
-~ 0.202468 {\cal P}_{\psi\psi\, (61)}
+ 0.607403 {\cal P}_{\psi\psi\, (62)}
- 0.054181 {\cal P}_{\psi\psi\, (63)}
\right. \nonumber \\
&& \left. ~
-~ 0.073883 {\cal P}_{\psi\psi\, (64)}
+ 0.123173 {\cal P}_{\psi\psi\, (65)}
- 0.369519 {\cal P}_{\psi\psi\, (66)}
\right. \nonumber \\
&& \left. ~
+~ 0.123173 {\cal P}_{\psi\psi\, (67)}
- 0.369519 {\cal P}_{\psi\psi\, (68)}
- 0.607403 {\cal P}_{\psi\psi\, (69)}
\right. \nonumber \\
&& \left. ~
+~ 0.202468 {\cal P}_{\psi\psi\, (70)}
- 0.369519 {\cal P}_{\psi\psi\, (71)}
+ 0.123173 {\cal P}_{\psi\psi\, (72)}
\right. \nonumber \\
&& \left. ~
-~ 0.369519 {\cal P}_{\psi\psi\, (73)}
+ 0.123173 {\cal P}_{\psi\psi\, (74)}
- 0.073883 {\cal P}_{\psi\psi\, (75)}
\right. \nonumber \\
&& \left. ~
-~ 0.054181 {\cal P}_{\psi\psi\, (76)}
- 0.073883 {\cal P}_{\psi\psi\, (77)}
- 0.054181 {\cal P}_{\psi\psi\, (78)}
\right. \nonumber \\
&& \left. ~
-~ 0.073883 {\cal P}_{\psi\psi\, (79)}
- 0.054181 {\cal P}_{\psi\psi\, (80)}
- 0.369519 {\cal P}_{\psi\psi\, (81)}
\right. \nonumber \\
&& \left. ~
+~ 0.123173 {\cal P}_{\psi\psi\, (82)}
+ 0.123173 {\cal P}_{\psi\psi\, (83)}
- 0.369519 {\cal P}_{\psi\psi\, (84)}
\right. \nonumber \\
&& \left. ~
-~ 0.054181 {\cal P}_{\psi\psi\, (85)}
- 0.073883 {\cal P}_{\psi\psi\, (86)}
- 0.054181 {\cal P}_{\psi\psi\, (87)}
\right. \nonumber \\
&& \left. ~
-~ 0.073883 {\cal P}_{\psi\psi\, (88)}
- 0.023225 {\cal P}_{\psi\psi\, (89)}
- 0.031671 {\cal P}_{\psi\psi\, (90)}
\right. \nonumber \\
&& \left. ~
+~ 0.031671 {\cal P}_{\psi\psi\, (91)}
+ 0.023225 {\cal P}_{\psi\psi\, (92)}
- 0.043943 {\cal P}_{\psi\psi\, (93)}
\right. \nonumber \\
&& \left. ~
-~ 0.059922 {\cal P}_{\psi\psi\, (94)}
+ 0.064498 {\cal P}_{\psi\psi\, (95)}
- 0.241431 {\cal P}_{\psi\psi\, (96)}
\right. \nonumber \\
&& \left. ~
+~ 0.086469 {\cal P}_{\psi\psi\, (97)}
- 0.211470 {\cal P}_{\psi\psi\, (98)}
+ 0.059922 {\cal P}_{\psi\psi\, (99)}
\right. \nonumber \\
&& \left. ~
+~ 0.043943 {\cal P}_{\psi\psi\, (100)}
+ 0.029961 {\cal P}_{\psi\psi\, (101)}
+ 0.021972 {\cal P}_{\psi\psi\, (102)}
\right. \nonumber \\
&& \left. ~
+~ 0.011984 {\cal P}_{\psi\psi\, (103)}
- 0.019974 {\cal P}_{\psi\psi\, (104)}
+ 0.029961 {\cal P}_{\psi\psi\, (105)}
\right. \nonumber \\
&& \left. ~
+~ 0.021972 {\cal P}_{\psi\psi\, (106)}
+ 0.211470 {\cal P}_{\psi\psi\, (107)}
- 0.086469 {\cal P}_{\psi\psi\, (108)}
\right. \nonumber \\
&& \left. ~
-~ 0.001997 {\cal P}_{\psi\psi\, (109)}
- 0.041946 {\cal P}_{\psi\psi\, (110)}
+ 0.019974 {\cal P}_{\psi\psi\, (111)}
\right. \nonumber \\
&& \left. ~
-~ 0.011984 {\cal P}_{\psi\psi\, (112)}
- 0.043943 {\cal P}_{\psi\psi\, (113)}
- 0.059922 {\cal P}_{\psi\psi\, (114)}
\right. \nonumber \\
&& \left. ~
-~ 0.021972 {\cal P}_{\psi\psi\, (115)}
- 0.029961 {\cal P}_{\psi\psi\, (116)}
+ 0.059922 {\cal P}_{\psi\psi\, (117)}
\right. \nonumber \\
&& \left. ~
+~ 0.043943 {\cal P}_{\psi\psi\, (118)}
+ 0.041946 {\cal P}_{\psi\psi\, (119)}
+ 0.001997 {\cal P}_{\psi\psi\, (120)}
\right. \nonumber \\
&& \left. ~
+~ 0.241431 {\cal P}_{\psi\psi\, (121)}
- 0.064498 {\cal P}_{\psi\psi\, (122)}
- 0.021972 {\cal P}_{\psi\psi\, (123)}
\right. \nonumber \\
&& \left. ~
-~ 0.029961 {\cal P}_{\psi\psi\, (124)}
- 0.087489 {\cal P}_{\psi\psi\, (125)}
+ 0.052717 {\cal P}_{\psi\psi\, (126)}
\right. \nonumber \\
&& \left. ~
-~ 0.041477 {\cal P}_{\psi\psi\, (127)}
+ 0.073977 {\cal P}_{\psi\psi\, (128)}
- 0.105841 {\cal P}_{\psi\psi\, (129)}
\right. \nonumber \\
&& \left. ~
+~ 0.158228 {\cal P}_{\psi\psi\, (130)}
+ 0.099560 {\cal P}_{\psi\psi\, (131)}
+ 0.073010 {\cal P}_{\psi\psi\, (132)}
\right. \nonumber \\
&& \left. ~
+~ 0.078300 {\cal P}_{\psi\psi\, (133)}
+ 0.026998 {\cal P}_{\psi\psi\, (134)}
+ 0.021259 {\cal P}_{\psi\psi\, (135)}
\right. \nonumber \\
&& \left. ~
+~ 0.046012 {\cal P}_{\psi\psi\, (136)}
+ 0.257788 {\cal P}_{\psi\psi\, (137)}
- 0.032831 {\cal P}_{\psi\psi\, (138)}
\right. \nonumber \\
&& \left. ~
-~ 0.158228 {\cal P}_{\psi\psi\, (139)}
+ 0.105841 {\cal P}_{\psi\psi\, (140)}
+ 0.018352 {\cal P}_{\psi\psi\, (141)}
\right. \nonumber \\
&& \left. ~
-~ 0.105511 {\cal P}_{\psi\psi\, (142)}
- 0.046012 {\cal P}_{\psi\psi\, (143)}
- 0.021259 {\cal P}_{\psi\psi\, (144)}
\right. \nonumber \\
&& \left. ~
-~ 0.043943 {\cal P}_{\psi\psi\, (145)}
- 0.059922 {\cal P}_{\psi\psi\, (146)}
+ 0.059922 {\cal P}_{\psi\psi\, (147)}
\right. \nonumber \\
&& \left. ~
+~ 0.043943 {\cal P}_{\psi\psi\, (148)}
- 0.008646 {\cal P}_{\psi\psi\, (149)}
- 0.183811 {\cal P}_{\psi\psi\, (150)}
\right. \nonumber \\
&& \left. ~
+~ 0.032831 {\cal P}_{\psi\psi\, (151)}
- 0.257788 {\cal P}_{\psi\psi\, (152)}
+ 0.183811 {\cal P}_{\psi\psi\, (153)}
\right. \nonumber \\
&& \left. ~
+~ 0.008646 {\cal P}_{\psi\psi\, (154)}
+ 0.105511 {\cal P}_{\psi\psi\, (155)}
- 0.018352 {\cal P}_{\psi\psi\, (156)}
\right. \nonumber \\
&& \left. ~
-~ 0.073977 {\cal P}_{\psi\psi\, (157)}
+ 0.041477 {\cal P}_{\psi\psi\, (158)}
- 0.026998 {\cal P}_{\psi\psi\, (159)}
\right. \nonumber \\
&& \left. ~
-~ 0.078300 {\cal P}_{\psi\psi\, (160)}
- 0.073010 {\cal P}_{\psi\psi\, (161)}
- 0.099560 {\cal P}_{\psi\psi\, (162)}
\right. \nonumber \\
&& \left. ~
-~ 0.052717 {\cal P}_{\psi\psi\, (163)}
+ 0.087489 {\cal P}_{\psi\psi\, (164)}
- 0.080757 {\cal P}_{\psi\psi\, (165)}
\right. \nonumber \\
&& \left. ~
-~ 0.110123 {\cal P}_{\psi\psi\, (166)}
+ 0.110123 {\cal P}_{\psi\psi\, (167)}
+ 0.080757 {\cal P}_{\psi\psi\, (168)} \right] a \nonumber \\
&& +~ O(a^2) 
\label{qcdquark}
\end{eqnarray}
where the colour group dependence is present in each tensor basis element. 
Again there are remnants of the symmetric point present in the expression.
However as this was a direct evaluation it is not clear if all colour channels
have a contribution for instance. Beyond one loop, by contrast, we would expect
additional Lorentz structures with respect to our use of the 
$\Gamma_{(n)}$-matrices similar to \cite{26,27}. 

\sect{QED results.}

Having concentrated on QCD we now turn to the case of QED and record the
analogous results for the quartic photon, photon-electron and quartic electron
Green's functions. The resulting expressions are not obtained by taking a group
theory limit but instead are a direct evaluation. Again for space reasons we 
present the results numerically in the Landau gauge. First, for the photon
$4$-point function we have 
\begin{eqnarray}
\left. \Sigma^{\QEDs}_{AA\, \mu \nu \sigma \rho}(p,q,r)
\right|_{\mbox{\footnotesize{symm}}} &=& 
\left[ 
-~ 0.716260 {\cal P}_{AA\, (1)} 
- 0.716260 {\cal P}_{AA\, (2)} 
- 0.716260 {\cal P}_{AA\, (3)} 
\right. \nonumber \\
&& \left. ~
- 2.382957 {\cal P}_{AA\, (4)} 
- 4.268597 {\cal P}_{AA\, (5)} 
+ 4.502776 {\cal P}_{AA\, (6)} 
\right. \nonumber \\
&& \left. ~
- 4.268597 {\cal P}_{AA\, (7)} 
- 2.382957 {\cal P}_{AA\, (8)} 
+ 4.502776 {\cal P}_{AA\, (9)} 
\right. \nonumber \\
&& \left. ~
- 2.217185 {\cal P}_{AA\, (10)} 
- 2.217185 {\cal P}_{AA\, (11)} 
+ 2.285591 {\cal P}_{AA\, (12)} 
\right. \nonumber \\
&& \left. ~
- 2.382957 {\cal P}_{AA\, (13)} 
+ 4.502776 {\cal P}_{AA\, (14)} 
- 4.268597 {\cal P}_{AA\, (15)} 
\right. \nonumber \\
&& \left. ~
- 2.217185 {\cal P}_{AA\, (16)} 
+ 2.285591 {\cal P}_{AA\, (17)} 
- 2.217185 {\cal P}_{AA\, (18)} 
\right. \nonumber \\
&& \left. ~
- 4.268597 {\cal P}_{AA\, (19)} 
+ 4.502776 {\cal P}_{AA\, (20)} 
- 2.382957 {\cal P}_{AA\, (21)} 
\right. \nonumber \\
&& \left. ~
+ 3.771280 {\cal P}_{AA\, (22)} 
+ 1.885640 {\cal P}_{AA\, (23)} 
+ 1.885640 {\cal P}_{AA\, (24)} 
\right. \nonumber \\
&& \left. ~
+ 1.885640 {\cal P}_{AA\, (25)} 
- 2.382957 {\cal P}_{AA\, (26)} 
- 6.885733 {\cal P}_{AA\, (27)} 
\right. \nonumber \\
&& \left. ~
+ 1.885640 {\cal P}_{AA\, (28)} 
- 0.165772 {\cal P}_{AA\, (29)} 
- 2.382957 {\cal P}_{AA\, (30)} 
\right. \nonumber \\
&& \left. ~
+ 2.285591 {\cal P}_{AA\, (31)} 
- 2.217185 {\cal P}_{AA\, (32)} 
- 2.217185 {\cal P}_{AA\, (33)} 
\right. \nonumber \\
&& \left. ~
+ 4.502776 {\cal P}_{AA\, (34)} 
- 2.382957 {\cal P}_{AA\, (35)} 
- 4.268597 {\cal P}_{AA\, (36)} 
\right. \nonumber \\
&& \left. ~
+ 4.502776 {\cal P}_{AA\, (37)} 
- 4.268597 {\cal P}_{AA\, (38)} 
- 2.382957 {\cal P}_{AA\, (39)} 
\right. \nonumber \\
&& \left. ~
- 2.382957 {\cal P}_{AA\, (40)} 
+ 1.885640 {\cal P}_{AA\, (41)} 
- 0.165772 {\cal P}_{AA\, (42)} 
\right. \nonumber \\
&& \left. ~
+ 1.885640 {\cal P}_{AA\, (43)} 
+ 3.771280 {\cal P}_{AA\, (44)} 
+ 1.885640 {\cal P}_{AA\, (45)} 
\right. \nonumber \\
&& \left. ~
- 6.885733 {\cal P}_{AA\, (46)} 
+ 1.885640 {\cal P}_{AA\, (47)} 
- 2.382957 {\cal P}_{AA\, (48)} 
\right. \nonumber \\
&& \left. ~
- 2.382957 {\cal P}_{AA\, (49)} 
- 0.165772 {\cal P}_{AA\, (50)} 
+ 1.885640 {\cal P}_{AA\, (51)} 
\right. \nonumber \\
&& \left. ~
- 6.885733 {\cal P}_{AA\, (52)} 
- 2.382957 {\cal P}_{AA\, (53)} 
+ 1.885640 {\cal P}_{AA\, (54)} 
\right. \nonumber \\
&& \left. ~
+ 1.885640 {\cal P}_{AA\, (55)} 
+ 1.885640 {\cal P}_{AA\, (56)} 
+ 3.771280 {\cal P}_{AA\, (57)} 
\right. \nonumber \\
&& \left. ~
- 1.307807 {\cal P}_{AA\, (58)} 
- 1.307807 {\cal P}_{AA\, (59)} 
- 1.307807 {\cal P}_{AA\, (60)} 
\right. \nonumber \\
&& \left. ~
- 0.838049 {\cal P}_{AA\, (61)} 
- 0.653903 {\cal P}_{AA\, (62)} 
- 0.653903 {\cal P}_{AA\, (63)} 
\right. \nonumber \\
&& \left. ~
- 0.469758 {\cal P}_{AA\, (64)} 
- 0.838049 {\cal P}_{AA\, (65)} 
- 0.653903 {\cal P}_{AA\, (66)} 
\right. \nonumber \\
&& \left. ~
- 0.653903 {\cal P}_{AA\, (67)} 
- 0.469758 {\cal P}_{AA\, (68)} 
- 0.838049 {\cal P}_{AA\, (69)} 
\right. \nonumber \\
&& \left. ~
- 0.653903 {\cal P}_{AA\, (70)} 
- 0.469758 {\cal P}_{AA\, (71)} 
- 0.653903 {\cal P}_{AA\, (72)} 
\right. \nonumber \\
&& \left. ~
- 0.838049 {\cal P}_{AA\, (73)} 
- 0.469758 {\cal P}_{AA\, (74)} 
- 0.653903 {\cal P}_{AA\, (75)} 
\right. \nonumber \\
&& \left. ~
- 0.653903 {\cal P}_{AA\, (76)} 
- 0.838049 {\cal P}_{AA\, (77)} 
- 0.469758 {\cal P}_{AA\, (78)} 
\right. \nonumber \\
&& \left. ~
- 0.653903 {\cal P}_{AA\, (79)} 
- 0.653903 {\cal P}_{AA\, (80)} 
- 0.838049 {\cal P}_{AA\, (81)} 
\right. \nonumber \\
&& \left. ~
- 0.653903 {\cal P}_{AA\, (82)} 
- 0.469758 {\cal P}_{AA\, (83)} 
- 0.653903 {\cal P}_{AA\, (84)} 
\right. \nonumber \\
&& \left. ~
- 0.653903 {\cal P}_{AA\, (85)} 
- 0.838049 {\cal P}_{AA\, (86)} 
- 0.469758 {\cal P}_{AA\, (87)} 
\right. \nonumber \\
&& \left. ~
- 0.838049 {\cal P}_{AA\, (88)} 
- 0.469758 {\cal P}_{AA\, (89)} 
- 0.653903 {\cal P}_{AA\, (90)} 
\right. \nonumber \\
&& \left. ~
- 0.838049 {\cal P}_{AA\, (91)} 
- 0.653903 {\cal P}_{AA\, (92)} 
- 0.469758 {\cal P}_{AA\, (93)} 
\right. \nonumber \\
&& \left. ~
- 0.838049 {\cal P}_{AA\, (94)} 
- 0.653903 {\cal P}_{AA\, (95)} 
- 0.469758 {\cal P}_{AA\, (96)} 
\right. \nonumber \\
&& \left. ~
- 0.838049 {\cal P}_{AA\, (97)} 
- 0.469758 {\cal P}_{AA\, (98)} 
- 0.653903 {\cal P}_{AA\, (99)} 
\right. \nonumber \\
&& \left. ~
- 0.653903 {\cal P}_{AA\, (100)} 
- 0.838049 {\cal P}_{AA\, (101)} 
- 0.469758 {\cal P}_{AA\, (102)} 
\right. \nonumber \\
&& \left. ~
+ 5.472775 {\cal P}_{AA\, (103)} 
- 0.497317 {\cal P}_{AA\, (104)} 
- 2.746924 {\cal P}_{AA\, (105)} 
\right. \nonumber \\
&& \left. ~
- 0.497317 {\cal P}_{AA\, (106)} 
- 0.313172 {\cal P}_{AA\, (107)} 
+ 5.656920 {\cal P}_{AA\, (108)} 
\right. \nonumber \\
&& \left. ~
- 0.497317 {\cal P}_{AA\, (109)} 
+ 5.472775 {\cal P}_{AA\, (110)} 
- 2.746924 {\cal P}_{AA\, (111)} 
\right. \nonumber \\
&& \left. ~
- 8.717016 {\cal P}_{AA\, (112)} 
+ 5.656920 {\cal P}_{AA\, (113)} 
- 0.313172 {\cal P}_{AA\, (114)} 
\right. \nonumber \\
&& \left. ~
- 0.497317 {\cal P}_{AA\, (115)} 
- 2.746924 {\cal P}_{AA\, (116)} 
+ 5.472775 {\cal P}_{AA\, (117)} 
\right. \nonumber \\
&& \left. ~
+ 5.656920 {\cal P}_{AA\, (118)} 
- 8.717016 {\cal P}_{AA\, (119)} 
+ 5.656920 {\cal P}_{AA\, (120)} 
\right. \nonumber \\
&& \left. ~
+ 5.472775 {\cal P}_{AA\, (121)} 
- 2.746924 {\cal P}_{AA\, (122)} 
- 0.497317 {\cal P}_{AA\, (123)} 
\right. \nonumber \\
&& \left. ~
- 0.313172 {\cal P}_{AA\, (124)} 
- 0.497317 {\cal P}_{AA\, (125)} 
- 0.313172 {\cal P}_{AA\, (126)} 
\right. \nonumber \\
&& \left. ~
- 2.746924 {\cal P}_{AA\, (127)} 
+ 5.472775 {\cal P}_{AA\, (128)} 
- 0.497317 {\cal P}_{AA\, (129)} 
\right. \nonumber \\
&& \left. ~
+ 5.656920 {\cal P}_{AA\, (130)} 
- 0.313172 {\cal P}_{AA\, (131)} 
- 0.497317 {\cal P}_{AA\, (132)} 
\right. \nonumber \\
&& \left. ~
- 2.746924 {\cal P}_{AA\, (133)} 
- 0.497317 {\cal P}_{AA\, (134)} 
+ 5.472775 {\cal P}_{AA\, (135)} 
\right. \nonumber \\
&& \left. ~
- 0.313172 {\cal P}_{AA\, (136)} 
+ 5.656920 {\cal P}_{AA\, (137)} 
- 8.717016 {\cal P}_{AA\, (138)} 
\right] \Nf a \nonumber \\
&& +~ O(a^2) 
\label{qedphot}
\end{eqnarray}
which involves all the possible Lorentz tensor structures which appear in the
quartic gluon Green's function of \cite{9}. Unlike the quartic gluon Green's
function (\ref{qedphot}) begins at $O(a)$ where in QED $a$ is related to the
fine structure constant as opposed to the strong coupling constant. The one 
loop expression is proportional to the number of massless electrons as the only
graphs which contribute at this order are closed electron boxes. The situation 
for the photon-electron $4$-point function has parallels to 
(\ref{qedphot}) as 
\begin{eqnarray}
\left. \Sigma^{\QEDs}_{A\psi\,\sigma \rho}(p,q,r)
\right|_{\mbox{\footnotesize{symm}}} &=& 
\left[ 
  0.368797 {\cal P}_{A\psi\, (1)} 
- 0.368797 {\cal P}_{A\psi\, (2)} 
+ 0.368797 {\cal P}_{A\psi\, (4)} 
\right. \nonumber \\
&& \left. ~
- 0.368797 {\cal P}_{A\psi\, (5)} 
- 0.406427 {\cal P}_{A\psi\, (7)} 
+ 0.406427 {\cal P}_{A\psi\, (8)} 
\right. \nonumber \\
&& \left. ~
+ 0.259319 {\cal P}_{A\psi\, (10)} 
+ 0.590103 {\cal P}_{A\psi\, (11)} 
+ 0.165609 {\cal P}_{A\psi\, (12)} 
\right. \nonumber \\
&& \left. ~
- 0.978151 {\cal P}_{A\psi\, (13)} 
- 1.099679 {\cal P}_{A\psi\, (14)} 
- 0.247840 {\cal P}_{A\psi\, (15)} 
\right. \nonumber \\
&& \left. ~
- 0.248659 {\cal P}_{A\psi\, (16)} 
- 0.286135 {\cal P}_{A\psi\, (17)} 
- 0.308512 {\cal P}_{A\psi\, (18)} 
\right. \nonumber \\
&& \left. ~
+ 1.099679 {\cal P}_{A\psi\, (19)} 
+ 0.978151 {\cal P}_{A\psi\, (20)} 
+ 0.247840 {\cal P}_{A\psi\, (21)} 
\right. \nonumber \\
&& \left. ~
- 0.590103 {\cal P}_{A\psi\, (22)} 
- 0.259319 {\cal P}_{A\psi\, (23)} 
- 0.165609 {\cal P}_{A\psi\, (24)} 
\right. \nonumber \\
&& \left. ~
+ 0.286135 {\cal P}_{A\psi\, (25)} 
+ 0.248659 {\cal P}_{A\psi\, (26)} 
+ 0.308512 {\cal P}_{A\psi\, (27)} 
\right. \nonumber \\
&& \left. ~
+ 0.225463 {\cal P}_{A\psi\, (28)} 
+ 1.380267 {\cal P}_{A\psi\, (29)} 
+ 0.225463 {\cal P}_{A\psi\, (30)} 
\right. \nonumber \\
&& \left. ~
- 1.380267 {\cal P}_{A\psi\, (31)} 
- 0.225463 {\cal P}_{A\psi\, (32)} 
- 0.225463 {\cal P}_{A\psi\, (33)} 
\right. \nonumber \\
&& \left. ~
+ 0.225463 {\cal P}_{A\psi\, (34)} 
- 0.225463 {\cal P}_{A\psi\, (35)} 
- 0.645223 {\cal P}_{A\psi\, (37)} 
\right. \nonumber \\
&& \left. ~
- 0.645223 {\cal P}_{A\psi\, (38)} 
- 1.290446 {\cal P}_{A\psi\, (39)} 
- 0.595466 {\cal P}_{A\psi\, (40)} 
\right. \nonumber \\
&& \left. ~
+ 0.595466 {\cal P}_{A\psi\, (41)} 
- 2.486948 {\cal P}_{A\psi\, (43)} 
- 2.603035 {\cal P}_{A\psi\, (44)} 
\right. \nonumber \\
&& \left. ~
- 0.977477 {\cal P}_{A\psi\, (45)} 
+ 2.603035 {\cal P}_{A\psi\, (46)} 
+ 2.486948 {\cal P}_{A\psi\, (47)} 
\right. \nonumber \\
&& \left. ~
+ 0.977477 {\cal P}_{A\psi\, (48)} 
- 0.479380 {\cal P}_{A\psi\, (49)} 
+ 0.479380 {\cal P}_{A\psi\, (50)} 
\right. \nonumber \\
&& \left. ~
+ 1.509471 {\cal P}_{A\psi\, (52)} 
+ 1.625558 {\cal P}_{A\psi\, (53)} 
- 0.977477 {\cal P}_{A\psi\, (54)} 
\right. \nonumber \\
&& \left. ~
- 1.625558 {\cal P}_{A\psi\, (55)} 
- 1.509471 {\cal P}_{A\psi\, (56)} 
+ 0.977477 {\cal P}_{A\psi\, (57)} 
\right. \nonumber \\
&& \left. ~
- 0.461758 {\cal P}_{A\psi\, (59)} 
- 0.862883 {\cal P}_{A\psi\, (60)} 
- 0.461758 {\cal P}_{A\psi\, (61)} 
\right. \nonumber \\
&& \left. ~
+ 0.862883 {\cal P}_{A\psi\, (62)} 
+ 0.461758 {\cal P}_{A\psi\, (63)} 
+ 0.461758 {\cal P}_{A\psi\, (64)} 
\right. \nonumber \\
&& \left. ~
- 0.461758 {\cal P}_{A\psi\, (65)} 
+ 0.461758 {\cal P}_{A\psi\, (66)} 
- 3.829066 {\cal P}_{A\psi\, (68)} 
\right] a \nonumber \\
&& +~ O(a^2) ~.
\end{eqnarray}
However, unlike (\ref{qedphot}) not all the Lorentz tensor structures which are
in the gluon-quark basis are used in this Green's function. What is more
evident in these two Green's functions is that there are more common 
coefficients of various tensors unlike the non-abelian case. 

Finally, the situation of the quartic electron Green's function is somewhat
different from its non-abelian counterpart. This is because of the way we had
to perform the computation due to the two independent spinor lines. As a 
consequence of this way we had to organize the calculation we have not used the
same Lorentz tensor basis as (\ref{qcdquark}). Instead we have constructed a
set of tensors specific to this electron Green's function and the explicit
forms are given in the Appendix. The full expression is 
\begin{eqnarray}
\left. \Sigma^{\QEDs}_{\psi\psi} (p,q,r) \right|_{\mbox{\footnotesize{symm}}}
&=& 
\left[ 
-~ 0.413930 {\cal P}^{\QEDs}_{\psi\psi\, (1)} 
+ 1.365234 {\cal P}^{\QEDs}_{\psi\psi\, (2)} 
+ 0.951304 {\cal P}^{\QEDs}_{\psi\psi\, (3)} 
\right. \nonumber \\
&& \left. ~
- 0.118477 {\cal P}^{\QEDs}_{\psi\psi\, (4)} 
- 1.779164 {\cal P}^{\QEDs}_{\psi\psi\, (5)} 
+ 1.779164 {\cal P}^{\QEDs}_{\psi\psi\, (6)} 
\right. \nonumber \\
&& \left. ~
- 0.236954 {\cal P}^{\QEDs}_{\psi\psi\, (7)} 
+ 0.951304 {\cal P}^{\QEDs}_{\psi\psi\, (8)} 
- 0.951304 {\cal P}^{\QEDs}_{\psi\psi\, (9)} 
\right. \nonumber \\
&& \left. ~
- 1.365234 {\cal P}^{\QEDs}_{\psi\psi\, (10)} 
- 1.779164 {\cal P}^{\QEDs}_{\psi\psi\, (11)} 
- 0.118477 {\cal P}^{\QEDs}_{\psi\psi\, (12)} 
\right. \nonumber \\
&& \left. ~
+ 0.413930 {\cal P}^{\QEDs}_{\psi\psi\, (13)} 
- 0.951304 {\cal P}^{\QEDs}_{\psi\psi\, (14)} 
- 0.118477 {\cal P}^{\QEDs}_{\psi\psi\, (15)} 
\right. \nonumber \\
&& \left. ~
- 0.413930 {\cal P}^{\QEDs}_{\psi\psi\, (16)} 
+ 1.365234 {\cal P}^{\QEDs}_{\psi\psi\, (17)} 
- 0.118477 {\cal P}^{\QEDs}_{\psi\psi\, (18)} 
\right. \nonumber \\
&& \left. ~
- 1.365234 {\cal P}^{\QEDs}_{\psi\psi\, (19)} 
+ 0.118477 {\cal P}^{\QEDs}_{\psi\psi\, (20)} 
+ 1.779164 {\cal P}^{\QEDs}_{\psi\psi\, (21)} 
\right. \nonumber \\
&& \left. ~
+ 0.118477 {\cal P}^{\QEDs}_{\psi\psi\, (22)} 
- 0.236954 {\cal P}^{\QEDs}_{\psi\psi\, (23)} 
+ 0.413930 {\cal P}^{\QEDs}_{\psi\psi\, (24)} 
\right. \nonumber \\
&& \left. ~
- 0.118477 {\cal P}^{\QEDs}_{\psi\psi\, (25)} 
- 0.118477 {\cal P}^{\QEDs}_{\psi\psi\, (26)} 
- 0.354640 {\cal P}^{\QEDs}_{\psi\psi\, (27)} 
\right. \nonumber \\
&& \left. ~
- 0.354640 {\cal P}^{\QEDs}_{\psi\psi\, (28)} 
- 0.354640 {\cal P}^{\QEDs}_{\psi\psi\, (29)} 
- 0.354640 {\cal P}^{\QEDs}_{\psi\psi\, (30)} 
\right. \nonumber \\
&& \left. ~
- 0.354640 {\cal P}^{\QEDs}_{\psi\psi\, (31)} 
- 0.354640 {\cal P}^{\QEDs}_{\psi\psi\, (32)} 
- 0.152021 {\cal P}^{\QEDs}_{\psi\psi\, (33)} 
\right. \nonumber \\
&& \left. ~
+ 0.152021 {\cal P}^{\QEDs}_{\psi\psi\, (34)} 
- 0.287628 {\cal P}^{\QEDs}_{\psi\psi\, (35)} 
- 0.071907 {\cal P}^{\QEDs}_{\psi\psi\, (36)} 
\right. \nonumber \\
&& \left. ~
+ 0.071907 {\cal P}^{\QEDs}_{\psi\psi\, (37)} 
+ 0.287628 {\cal P}^{\QEDs}_{\psi\psi\, (38)} 
+ 0.14381 {\cal P}^{\QEDs}_{\psi\psi\, (39)} 
\right. \nonumber \\
&& \left. ~
- 0.071907 {\cal P}^{\QEDs}_{\psi\psi\, (40)} 
+ 0.143814 {\cal P}^{\QEDs}_{\psi\psi\, (41)} 
- 0.071907 {\cal P}^{\QEDs}_{\psi\psi\, (42)} 
\right. \nonumber \\
&& \left. ~
- 0.071907 {\cal P}^{\QEDs}_{\psi\psi\, (43)} 
+ 0.071907 {\cal P}^{\QEDs}_{\psi\psi\, (44)} 
- 0.287628 {\cal P}^{\QEDs}_{\psi\psi\, (45)} 
\right. \nonumber \\
&& \left. ~
- 0.143814 {\cal P}^{\QEDs}_{\psi\psi\, (46)} 
+ 0.287628 {\cal P}^{\QEDs}_{\psi\psi\, (47)} 
+ 0.071907 {\cal P}^{\QEDs}_{\psi\psi\, (48)} 
\right. \nonumber \\
&& \left. ~
+ 0.071907 {\cal P}^{\QEDs}_{\psi\psi\, (49)} 
- 0.143814 {\cal P}^{\QEDs}_{\psi\psi\, (50)} 
- 0.314625 {\cal P}^{\QEDs}_{\psi\psi\, (51)} 
\right. \nonumber \\
&& \left. ~
- 0.075682 {\cal P}^{\QEDs}_{\psi\psi\, (52)} 
- 0.238943 {\cal P}^{\QEDs}_{\psi\psi\, (53)} 
+ 0.477887 {\cal P}^{\QEDs}_{\psi\psi\, (54)} 
\right. \nonumber \\
&& \left. ~
+ 0.238944 {\cal P}^{\QEDs}_{\psi\psi\, (55)} 
+ 0.238944 {\cal P}^{\QEDs}_{\psi\psi\, (56)} 
+ 0.238944 {\cal P}^{\QEDs}_{\psi\psi\, (57)} 
\right. \nonumber \\
&& \left. ~
+ 0.238944 {\cal P}^{\QEDs}_{\psi\psi\, (58)} 
- 0.075682 {\cal P}^{\QEDs}_{\psi\psi\, (59)} 
- 0.238943 {\cal P}^{\QEDs}_{\psi\psi\, (60)} 
\right. \nonumber \\
&& \left. ~
- 0.287628 {\cal P}^{\QEDs}_{\psi\psi\, (61)} 
+ 0.287628 {\cal P}^{\QEDs}_{\psi\psi\, (62)} 
- 0.314625 {\cal P}^{\QEDs}_{\psi\psi\, (63)} 
\right. \nonumber \\
&& \left. ~
- 0.238943 {\cal P}^{\QEDs}_{\psi\psi\, (64)} 
+ 0.314625 {\cal P}^{\QEDs}_{\psi\psi\, (65)} 
+ 0.075682 {\cal P}^{\QEDs}_{\psi\psi\, (66)} 
\right. \nonumber \\
&& \left. ~
+ 0.075682 {\cal P}^{\QEDs}_{\psi\psi\, (67)} 
- 0.238943 {\cal P}^{\QEDs}_{\psi\psi\, (68)} 
- 0.477887 {\cal P}^{\QEDs}_{\psi\psi\, (69)} 
\right. \nonumber \\
&& \left. ~
+ 0.314625 {\cal P}^{\QEDs}_{\psi\psi\, (70)} 
- 0.528590 {\cal P}^{\QEDs}_{\psi\psi\, (71)} 
+ 0.528590 {\cal P}^{\QEDs}_{\psi\psi\, (72)} 
\right] a \nonumber \\
&& +~ O(a^2) 
\end{eqnarray}
where alliances between tensor structures is again manifest. In contrast to
(\ref{qcdquark}) there are less than half the number of Lorentz structures.  

\sect{Discussion.}

We have completed the one loop evaluation of all the possible $4$-point Green's
functions in QCD in a linear covariant gauge at the completely symmetric 
subtraction point. The initial computations of \cite{8,9} concentrated on the 
quartic gluon function also at the fully symmetric point and this article 
completes the symmetric point programme. While the aim was partly to achieve 
this, the results should prove useful to Schwinger-Dyson studies of $4$-point 
functions. The solutions to such equations should overlap with the perturbative 
information given here and can be used to ensure that the approximations used
to truncate the tower of Green's function are in fact consistent with 
independent results. In turn such consistency should impinge upon the 
Schwinger-Dyson analyses of the $3$-point functions of QCD where effects from 
$4$-point kernels have been ignored in the first instance and models of 
swordfish diagram contributions are now being used. There are several 
directions in which the present work can be extended to further complement such
Schwinger-Dyson analyses. The first is to go the next order in perturbation 
theory for all six $4$-point functions. This is not a straightforward task as 
the two loop master integrals at the fully symmetric subtraction point, which 
would emerge from a {\sc Reduze} reduction, are not yet known. Once these are 
available then the next order can be completed. For the five Green's functions 
considered here this will involve a renormalization which in passing acts as a 
check on the one loop analysis. A second direction would be to repeat the one 
loop computation but at the fully off-shell point. While the one loop masters 
are already known from \cite{35} the reduction of the necessary scalar Feynman 
integrals lurking within each Green's functions would need to be constructed 
first. Aside from these two tasks one immediate application of the formalism 
recorded here is to the study of QCD gauge fixing in gauges other than the 
linear covariant one. We have noted two such gauges which are the Curci-Ferrari
one, \cite{18}, and the maximal abelian gauge, \cite{36,37,38}. Although these 
are nonlinear gauges they have been of interest due to their potential 
connection with colour confinement. Indeed the $3$-point vertex functions of 
QCD have been studied in these other gauges using Schwinger-Dyson techniques in
order to extract a point of view on gluon confinement which is consistent in 
different gauges and hence is of physical relevance. A side dish to such a 
programme is ensuring the consistent implementation of the Slavnov-Taylor 
identities in the running down to the infrared region. Some such identities 
involve several of the Green's functions evaluated here and so our analysis 
should prove useful for that programme.  

\vspace{1cm}
\noindent
{\bf Acknowledgements.} The author thanks Prof. J. Pawlowski for discussions
and the encouragement to carry out this work as well as Prof. A. Eichhorn and 
the ITP, University of Heidelberg for hospitality during the visit where it was
initiated. The work was carried out with the support of the STFC through the
Consolidated Grant ST/L000431/1.

\appendix

\sect{Tensor bases.}

For reference we provide the Lorentz tensor bases for each of the three Green's
functions where we used the projection method. The basis which emerged from the
explicit one loop computation of the quartic quark Green's function is also
included as well as the analogous basis for QED. On notation, where a 
generalized $\Gamma{(n)}$-matrix has a contraction with an external momentum we
use the convention that the vector appears in place of the contracting index. 
First, the ten basis elements for the gluon-ghost function are  
\begin{eqnarray}
{\cal P}_{Ac\,(1)}^{\sigma\rho} &=& \eta^{\sigma\rho} ~~,~~
{\cal P}_{Ac\,(2)}^{\sigma\rho} ~=~ \frac{p^\sigma p^\rho}{\mu^2} ~~,~~
{\cal P}_{Ac\,(3)}^{\sigma\rho} ~=~ \frac{p^\sigma q^\rho}{\mu^2} ~~,~~ 
{\cal P}_{Ac\,(4)}^{\sigma\rho} ~=~ \frac{p^\sigma r^\rho}{\mu^2} \nonumber \\ 
{\cal P}_{Ac\,(5)}^{\sigma\rho} &=& \frac{q^\sigma p^\rho}{\mu^2} ~~,~~
{\cal P}_{Ac\,(6)}^{\sigma\rho} ~=~ \frac{q^\sigma q^\rho}{\mu^2} ~~,~~
{\cal P}_{Ac\,(7)}^{\sigma\rho} ~=~ \frac{q^\sigma r^\rho}{\mu^2} ~~,~~
{\cal P}_{Ac\,(8)}^{\sigma\rho} ~=~ \frac{r^\sigma p^\rho}{\mu^2} \nonumber \\
{\cal P}_{Ac\,(9)}^{\sigma\rho} &=& \frac{r^\sigma q^\rho}{\mu^2} ~~,~~ 
{\cal P}_{Ac\,(10)}^{\sigma\rho} ~=~ \frac{r^\sigma r^\rho}{\mu^2} ~.~~~
\end{eqnarray}
We have included the mass scale $\mu$ to ensure that all the tensors have the
same mass dimension. The colour group factor is absent in the Lorentz basis
but appears explicitly in the results section. For the gluon-quark $4$-point 
function there are $68$ tensors but with the use of the $\Gamma_{(n)}$-matrices
this partitions into subspaces of rank $36$, $31$ and $1$. We have 
\begin{eqnarray}
{\cal P}_{A\psi\,(1)}^{\sigma\rho}(p,q,r) &=& \gamma^{\sigma} p^\rho ~~,~~  
{\cal P}_{A\psi\,(2)}^{\sigma\rho}(p,q,r) ~=~ \gamma^{\sigma} q^\rho ~~,~~  
{\cal P}_{A\psi\,(3)}^{\sigma\rho}(p,q,r) ~=~ \gamma^{\sigma} r^\rho 
\nonumber \\
{\cal P}_{A\psi\,(4)}^{\sigma\rho}(p,q,r) &=& \gamma^{\rho} p^\sigma ~~,~~ 
{\cal P}_{A\psi\,(5)}^{\sigma\rho}(p,q,r) ~=~ \gamma^{\rho} q^\sigma ~~,~~ 
{\cal P}_{A\psi\,(6)}^{\sigma\rho}(p,q,r) ~=~ \gamma^{\rho} r^\sigma 
\nonumber \\ 
{\cal P}_{A\psi\,(7)}^{\sigma\rho}(p,q,r) &=& \pslash \eta^{\sigma\rho} ~~,~~ 
{\cal P}_{A\psi\,(8)}^{\sigma\rho}(p,q,r) ~=~ \qslash \eta^{\sigma\rho} ~~,~~ 
{\cal P}_{A\psi\,(9)}^{\sigma\rho}(p,q,r) ~=~ \rslash \eta^{\sigma\rho} 
\nonumber \\ 
{\cal P}_{A\psi\,(10)}^{\sigma\rho}(p,q,r) &=& 
\frac{\pslash p^\sigma p^\rho}{\mu^2} ~~,~~ 
{\cal P}_{A\psi\,(11)}^{\sigma\rho}(p,q,r) ~=~ 
\frac{\pslash p^\sigma q^\rho}{\mu^2} ~~,~~ 
{\cal P}_{A\psi\,(12)}^{\sigma\rho}(p,q,r) ~=~ 
\frac{\pslash p^\sigma r^\rho}{\mu^2} \nonumber \\ 
{\cal P}_{A\psi\,(13)}^{\sigma\rho}(p,q,r) &=& 
\frac{\pslash q^\sigma p^\rho}{\mu^2} ~~,~~
{\cal P}_{A\psi\,(14)}^{\sigma\rho}(p,q,r) ~=~ 
\frac{\pslash q^\sigma q^\rho}{\mu^2} ~~,~~ 
{\cal P}_{A\psi\,(15)}^{\sigma\rho}(p,q,r) ~=~ 
\frac{\pslash q^\sigma r^\rho}{\mu^2} \nonumber \\ 
{\cal P}_{A\psi\,(16)}^{\sigma\rho}(p,q,r) &=& 
\frac{\pslash r^\sigma p^\rho}{\mu^2} ~~,~~ 
{\cal P}_{A\psi\,(17)}^{\sigma\rho}(p,q,r) ~=~ 
\frac{\pslash r^\sigma q^\rho}{\mu^2} ~~,~~ 
{\cal P}_{A\psi\,(18)}^{\sigma\rho}(p,q,r) ~=~ 
\frac{\pslash r^\sigma r^\rho}{\mu^2} \nonumber \\ 
{\cal P}_{A\psi\,(19)}^{\sigma\rho}(p,q,r) &=& 
\frac{\qslash p^\sigma p^\rho}{\mu^2} ~~,~~ 
{\cal P}_{A\psi\,(20)}^{\sigma\rho}(p,q,r) ~=~ 
\frac{\qslash p^\sigma q^\rho}{\mu^2} ~~,~~ 
{\cal P}_{A\psi\,(21)}^{\sigma\rho}(p,q,r) ~=~ 
\frac{\qslash p^\sigma r^\rho}{\mu^2} \nonumber \\ 
{\cal P}_{A\psi\,(22)}^{\sigma\rho}(p,q,r) &=& 
\frac{\qslash q^\sigma p^\rho}{\mu^2} ~~,~~ 
{\cal P}_{A\psi\,(23)}^{\sigma\rho}(p,q,r) ~=~ 
\frac{\qslash q^\sigma q^\rho}{\mu^2} ~~,~~ 
{\cal P}_{A\psi\,(24)}^{\sigma\rho}(p,q,r) ~=~ 
\frac{\qslash q^\sigma r^\rho}{\mu^2} \nonumber \\ 
{\cal P}_{A\psi\,(25)}^{\sigma\rho}(p,q,r) &=& 
\frac{\qslash r^\sigma p^\rho}{\mu^2} ~~,~~ 
{\cal P}_{A\psi\,(26)}^{\sigma\rho}(p,q,r) ~=~ 
\frac{\qslash r^\sigma q^\rho}{\mu^2} ~~,~~ 
{\cal P}_{A\psi\,(27)}^{\sigma\rho}(p,q,r) ~=~ 
\frac{\qslash r^\sigma r^\rho}{\mu^2} \nonumber \\ 
{\cal P}_{A\psi\,(28)}^{\sigma\rho}(p,q,r) &=& 
\frac{\rslash p^\sigma p^\rho}{\mu^2} ~~,~~ 
{\cal P}_{A\psi\,(29)}^{\sigma\rho}(p,q,r) ~=~ 
\frac{\rslash p^\sigma q^\rho}{\mu^2} ~~,~~ 
{\cal P}_{A\psi\,(30)}^{\sigma\rho}(p,q,r) ~=~ 
\frac{\rslash p^\sigma r^\rho}{\mu^2} \nonumber \\
{\cal P}_{A\psi\,(31)}^{\sigma\rho}(p,q,r) &=& 
\frac{\rslash q^\sigma p^\rho}{\mu^2} ~~,~~ 
{\cal P}_{A\psi\,(32)}^{\sigma\rho}(p,q,r) ~=~ 
\frac{\rslash q^\sigma q^\rho}{\mu^2} ~~,~~ 
{\cal P}_{A\psi\,(33)}^{\sigma\rho}(p,q,r) ~=~ 
\frac{\rslash q^\sigma r^\rho}{\mu^2} 
\nonumber \\
{\cal P}_{A\psi\,(34)}^{\sigma\rho}(p,q,r) &=& 
\frac{\rslash r^\sigma p^\rho}{\mu^2} ~~,~~ 
{\cal P}_{A\psi\,(35)}^{\sigma\rho}(p,q,r) ~=~ 
\frac{\rslash r^\sigma q^\rho}{\mu^2} ~~,~~ 
{\cal P}_{A\psi\,(36)}^{\sigma\rho}(p,q,r) ~=~ 
\frac{\rslash r^\sigma r^\rho}{\mu^2} \nonumber \\ 
{\cal P}_{A\psi\,(37)}^{\sigma\rho}(p,q,r) &=& 
\Gamma_{(3)}^{\sigma\rho p} ~~,~~ 
{\cal P}_{A\psi\,(38)}^{\sigma\rho}(p,q,r) ~=~ 
\Gamma_{(3)}^{\sigma\rho q} ~~,~~
{\cal P}_{A\psi\,(39)}^{\sigma\rho}(p,q,r) ~=~ 
\Gamma_{(3)}^{\sigma\rho r} \nonumber \\ 
{\cal P}_{A\psi\,(40)}^{\sigma\rho}(p,q,r) &=& 
\frac{\Gamma_{(3)}^{\sigma pq} p^\rho}{\mu^2} ~~,~~
{\cal P}_{A\psi\,(41)}^{\sigma\rho}(p,q,r) ~=~ 
\frac{\Gamma_{(3)}^{\sigma pq} q^\rho}{\mu^2} ~~,~~
{\cal P}_{A\psi\,(42)}^{\sigma\rho}(p,q,r) ~=~ 
\frac{\Gamma_{(3)}^{\sigma pq} r^\rho}{\mu^2} \nonumber \\ 
{\cal P}_{A\psi\,(43)}^{\sigma\rho}(p,q,r) &=& 
\frac{\Gamma_{(3)}^{\sigma qr} p^\rho}{\mu^2} ~~,~~
{\cal P}_{A\psi\,(44)}^{\sigma\rho}(p,q,r) ~=~ 
\frac{\Gamma_{(3)}^{\sigma qr} q^\rho}{\mu^2} ~~,~~
{\cal P}_{A\psi\,(45)}^{\sigma\rho}(p,q,r) ~=~ 
\frac{\Gamma_{(3)}^{\sigma qr} r^\rho}{\mu^2} \nonumber \\ 
{\cal P}_{A\psi\,(46)}^{\sigma\rho}(p,q,r) &=& 
\frac{\Gamma_{(3)}^{\sigma rp} p^\rho}{\mu^2} ~~,~~
{\cal P}_{A\psi\,(47)}^{\sigma\rho}(p,q,r) ~=~ 
\frac{\Gamma_{(3)}^{\sigma rp} q^\rho}{\mu^2} ~~,~~
{\cal P}_{A\psi\,(48)}^{\sigma\rho}(p,q,r) ~=~ 
\frac{\Gamma_{(3)}^{\sigma rp} r^\rho}{\mu^2} \nonumber \\ 
{\cal P}_{A\psi\,(49)}^{\sigma\rho}(p,q,r) &=& 
\frac{\Gamma_{(3)}^{\rho pq} p^\sigma}{\mu^2} ~~,~~
{\cal P}_{A\psi\,(50)}^{\sigma\rho}(p,q,r) ~=~ 
\frac{\Gamma_{(3)}^{\rho pq} q^\sigma}{\mu^2} ~~,~~
{\cal P}_{A\psi\,(51)}^{\sigma\rho}(p,q,r) ~=~ 
\frac{\Gamma_{(3)}^{\rho pq} r^\sigma}{\mu^2} \nonumber \\ 
{\cal P}_{A\psi\,(52)}^{\sigma\rho}(p,q,r) &=& 
\frac{\Gamma_{(3)}^{\rho qr} p^\sigma}{\mu^2} ~~,~~
{\cal P}_{A\psi\,(53)}^{\sigma\rho}(p,q,r) ~=~ 
\frac{\Gamma_{(3)}^{\rho qr} q^\sigma}{\mu^2} ~~,~~
{\cal P}_{A\psi\,(54)}^{\sigma\rho}(p,q,r) ~=~ 
\frac{\Gamma_{(3)}^{\rho qr} r^\sigma}{\mu^2} \nonumber \\ 
{\cal P}_{A\psi\,(55)}^{\sigma\rho}(p,q,r) &=& 
\frac{\Gamma_{(3)}^{\rho rp} p^\sigma}{\mu^2} ~~,~~
{\cal P}_{A\psi\,(56)}^{\sigma\rho}(p,q,r) ~=~ 
\frac{\Gamma_{(3)}^{\rho rp} q^\sigma}{\mu^2} ~~,~~
{\cal P}_{A\psi\,(57)}^{\sigma\rho}(p,q,r) ~=~ 
\frac{\Gamma_{(3)}^{\rho rp} r^\sigma}{\mu^2} \nonumber \\ 
{\cal P}_{A\psi\,(58)}^{\sigma\rho}(p,q,r) &=& \frac{\Gamma_{(3)}^{pqr} 
\eta^{\sigma\rho}}{\mu^2} ~~,~~ 
{\cal P}_{A\psi\,(59)}^{\sigma\rho}(p,q,r) ~=~ 
\frac{\Gamma_{(3)}^{pqr} p^\sigma p^\rho}{\mu^4} ~~,~~ 
{\cal P}_{A\psi\,(60)}^{\sigma\rho}(p,q,r) ~=~ 
\frac{\Gamma_{(3)}^{pqr} p^\sigma q^\rho}{\mu^4} \nonumber \\ 
{\cal P}_{A\psi\,(61)}^{\sigma\rho}(p,q,r) &=& 
\frac{\Gamma_{(3)}^{pqr} p^\sigma r^\rho}{\mu^4} ~~,~~ 
{\cal P}_{A\psi\,(62)}^{\sigma\rho}(p,q,r) ~=~ 
\frac{\Gamma_{(3)}^{pqr} q^\sigma p^\rho}{\mu^4} ~~,~~ 
{\cal P}_{A\psi\,(63)}^{\sigma\rho}(p,q,r) ~=~ 
\frac{\Gamma_{(3)}^{pqr} q^\sigma q^\rho}{\mu^4} \nonumber \\ 
{\cal P}_{A\psi\,(64)}^{\sigma\rho}(p,q,r) &=& 
\frac{\Gamma_{(3)}^{pqr} q^\sigma r^\rho}{\mu^4} ~~,~~ 
{\cal P}_{A\psi\,(65)}^{\sigma\rho}(p,q,r) ~=~ 
\frac{\Gamma_{(3)}^{pqr} r^\sigma p^\rho}{\mu^4} ~~,~~ 
{\cal P}_{A\psi\,(66)}^{\sigma\rho}(p,q,r) ~=~ 
\frac{\Gamma_{(3)}^{pqr} r^\sigma q^\rho}{\mu^4} \nonumber \\ 
{\cal P}_{A\psi\,(67)}^{\sigma\rho}(p,q,r) &=& 
\frac{\Gamma_{(3)}^{pqr} r^\sigma r^\rho}{\mu^4} ~~,~~ 
{\cal P}_{A\psi\,(68)}^{\sigma\rho}(p,q,r) ~=~ 
\frac{\Gamma_{(5)}^{\sigma\rho pqr}}{\mu^2} ~. 
\label{basisaq}
\end{eqnarray}
In the data file the projection matrix for this basis is presented in block
partition form. The elements of the $\Gamma_{(3)}$ partition there are 
numbered from $1$ to $31$ and that for $\Gamma_{(5)}$ is merely $1$. To map to
the labels for each sector to the full $68$~$\times$~$68$ projection matrix, 
for which (\ref{basisaq}) is the basis, $36$ needs to be added to the labels of
the $\Gamma_{(3)}$ partition and $67$ to that for the $\Gamma_{(5)}$ partition 
in order to produce all the {\em non-zero} entries in 
${\cal M}_{A\psi\, kk^\prime}$ where $1$~$\leq$~$k$~$\leq$~$68$. By contrast 
the basis for the ghost-quark $4$-point function is somewhat smaller as  
\begin{equation}
{\cal P}_{c\psi\,(1)} ~=~ \pslash ~~,~~
{\cal P}_{c\psi\,(2)} ~=~ \qslash ~~,~~
{\cal P}_{c\psi\,(3)} ~=~ \rslash ~~,~~
{\cal P}_{c\psi\,(4)} ~=~ \frac{\Gamma_{(3)}^{pqr}}{\mu^2} ~.
\end{equation}
It is worth noting that the basis involving external quark legs would be larger 
if there was a non-zero quark mass. 

As noted earlier the situation with the structure of the one loop quartic quark
$4$-point function is different. We have not endeavoured to construct the most
general basis of Lorentz tensors due to the separate spinor strings. Instead
we provide the tensors which emerged as a consequence of the direct {\em one}
loop computation to make contact with the electronic format of the data file. 
These are  
\begin{eqnarray}
{\cal P}_{\psi\psi\, (1)} &=& {\Gamma_{(1)}^p}_\alpha^{~\beta} {\Gamma_{(1)}^p}_\delta^{~\gamma} \delta^{ij} \delta^{kl} \delta_{IJ} \delta_{KL} 
~~~,~~~
{\cal P}_{\psi\psi\, (2)} ~=~ {\Gamma_{(1)}^p}_\alpha^{~\beta} {\Gamma_{(1)}^p}_\delta^{~\gamma} \delta^{ij} \delta^{kl} \delta_{IK} \delta_{JL} 
\nonumber \\
{\cal P}_{\psi\psi\, (3)} &=& {\Gamma_{(1)}^p}_\alpha^{~\beta} {\Gamma_{(1)}^q}_\delta^{~\gamma} \delta^{ij} \delta^{kl} \delta_{IJ} \delta_{KL} 
~~~,~~~ 
{\cal P}_{\psi\psi\, (4)} ~=~ {\Gamma_{(1)}^p}_\alpha^{~\beta} {\Gamma_{(1)}^q}_\delta^{~\gamma} \delta^{ij} \delta^{kl} \delta_{IK} \delta_{JL} 
\nonumber \\
{\cal P}_{\psi\psi\, (5)} &=& {\Gamma_{(1)}^p}_\alpha^{~\beta} {\Gamma_{(1)}^r}_\delta^{~\gamma} \delta^{ij} \delta^{kl} \delta_{IJ} \delta_{KL} 
~~~,~~~ 
{\cal P}_{\psi\psi\, (6)} ~=~ {\Gamma_{(1)}^p}_\alpha^{~\beta} {\Gamma_{(1)}^r}_\delta^{~\gamma} \delta^{ij} \delta^{kl} \delta_{IK} \delta_{JL} 
\nonumber \\
{\cal P}_{\psi\psi\, (7)} &=& {\Gamma_{(1)}^p}_\alpha^{~\beta} {\Gamma_{(3)}^{p q r}}_\delta^{~\gamma} \delta^{ij} \delta^{kl} \delta_{IJ} \delta_{KL}
~~~,~~~ 
{\cal P}_{\psi\psi\, (8)} ~=~ {\Gamma_{(1)}^p}_\alpha^{~\beta} {\Gamma_{(3)}^{p q r}}_\delta^{~\gamma} \delta^{ij} \delta^{kl} \delta_{IK} \delta_{JL}
\nonumber \\
{\cal P}_{\psi\psi\, (9)} &=& {\Gamma_{(1)}^p}_\alpha^{~\gamma} {\Gamma_{(1)}^p}_\delta^{~\beta} \delta^{ik} \delta^{jl} \delta_{IJ} \delta_{KL} 
~~~,~~~ 
{\cal P}_{\psi\psi\, (10)} ~=~ {\Gamma_{(1)}^p}_\alpha^{~\gamma} {\Gamma_{(1)}^p}_\delta^{~\beta} \delta^{ik} \delta^{jl} \delta_{IK} \delta_{JL} 
\nonumber \\
{\cal P}_{\psi\psi\, (11)} &=& {\Gamma_{(1)}^p}_\alpha^{~\gamma} {\Gamma_{(1)}^q}_\delta^{~\beta} \delta^{ik} \delta^{jl} \delta_{IJ} \delta_{KL} 
~~~,~~~ 
{\cal P}_{\psi\psi\, (12)} ~=~ {\Gamma_{(1)}^p}_\alpha^{~\gamma} {\Gamma_{(1)}^q}_\delta^{~\beta} \delta^{ik} \delta^{jl} \delta_{IK} \delta_{JL} 
\nonumber \\
{\cal P}_{\psi\psi\, (13)} &=& {\Gamma_{(1)}^p}_\alpha^{~\gamma} {\Gamma_{(1)}^r}_\delta^{~\beta} \delta^{ik} \delta^{jl} \delta_{IJ} \delta_{KL} 
~~~,~~~ 
{\cal P}_{\psi\psi\, (14)} ~=~ {\Gamma_{(1)}^p}_\alpha^{~\gamma} {\Gamma_{(1)}^r}_\delta^{~\beta} \delta^{ik} \delta^{jl} \delta_{IK} \delta_{JL} 
\nonumber \\
{\cal P}_{\psi\psi\, (15)} &=& {\Gamma_{(1)}^p}_\alpha^{~\gamma} {\Gamma_{(3)}^{p q r}}_\delta^{~\beta} \delta^{ik} \delta^{jl} \delta_{IJ} \delta_{KL}
~~~,~~~ 
{\cal P}_{\psi\psi\, (16)} ~=~ {\Gamma_{(1)}^p}_\alpha^{~\gamma} {\Gamma_{(3)}^{p q r}}_\delta^{~\beta} \delta^{ik} \delta^{jl} \delta_{IK} \delta_{JL}
\nonumber \\
{\cal P}_{\psi\psi\, (17)} &=& {\Gamma_{(1)}^p}_\delta^{~\beta} {\Gamma_{(1)}^q}_\alpha^{~\gamma} \delta^{ik} \delta^{jl} \delta_{IJ} \delta_{KL} 
~~~,~~~ 
{\cal P}_{\psi\psi\, (18)} ~=~ {\Gamma_{(1)}^p}_\delta^{~\beta} {\Gamma_{(1)}^q}_\alpha^{~\gamma} \delta^{ik} \delta^{jl} \delta_{IK} \delta_{JL} 
\nonumber \\
{\cal P}_{\psi\psi\, (19)} &=& {\Gamma_{(1)}^p}_\delta^{~\beta} {\Gamma_{(1)}^r}_\alpha^{~\gamma} \delta^{ik} \delta^{jl} \delta_{IJ} \delta_{KL} 
~~~,~~~ 
{\cal P}_{\psi\psi\, (20)} ~=~ {\Gamma_{(1)}^p}_\delta^{~\beta} {\Gamma_{(1)}^r}_\alpha^{~\gamma} \delta^{ik} \delta^{jl} \delta_{IK} \delta_{JL} 
\nonumber \\
{\cal P}_{\psi\psi\, (21)} &=& {\Gamma_{(1)}^p}_\delta^{~\beta} {\Gamma_{(3)}^{p q r}}_\alpha^{~\gamma} \delta^{ik} \delta^{jl} \delta_{IJ} \delta_{KL}
~~~,~~~ 
{\cal P}_{\psi\psi\, (22)} ~=~ {\Gamma_{(1)}^p}_\delta^{~\beta} {\Gamma_{(3)}^{p q r}}_\alpha^{~\gamma} \delta^{ik} \delta^{jl} \delta_{IK} \delta_{JL}
\nonumber \\
{\cal P}_{\psi\psi\, (23)} &=& {\Gamma_{(1)}^p}_\delta^{~\gamma} {\Gamma_{(1)}^q}_\alpha^{~\beta} \delta^{ij} \delta^{kl} \delta_{IJ} \delta_{KL} 
~~~,~~~ 
{\cal P}_{\psi\psi\, (24)} ~=~ {\Gamma_{(1)}^p}_\delta^{~\gamma} {\Gamma_{(1)}^q}_\alpha^{~\beta} \delta^{ij} \delta^{kl} \delta_{IK} \delta_{JL} 
\nonumber \\
{\cal P}_{\psi\psi\, (25)} &=& {\Gamma_{(1)}^p}_\delta^{~\gamma} {\Gamma_{(1)}^r}_\alpha^{~\beta} \delta^{ij} \delta^{kl} \delta_{IJ} \delta_{KL} 
~~~,~~~ 
{\cal P}_{\psi\psi\, (26)} ~=~ {\Gamma_{(1)}^p}_\delta^{~\gamma} {\Gamma_{(1)}^r}_\alpha^{~\beta} \delta^{ij} \delta^{kl} \delta_{IK} \delta_{JL} 
\nonumber \\
{\cal P}_{\psi\psi\, (27)} &=& {\Gamma_{(1)}^p}_\delta^{~\gamma} {\Gamma_{(3)}^{p q r}}_\alpha^{~\beta} \delta^{ij} \delta^{kl} \delta_{IJ} \delta_{KL}
~~~,~~~ 
{\cal P}_{\psi\psi\, (28)} ~=~ {\Gamma_{(1)}^p}_\delta^{~\gamma} {\Gamma_{(3)}^{p q r}}_\alpha^{~\beta} \delta^{ij} \delta^{kl} \delta_{IK} \delta_{JL}
\nonumber \\
{\cal P}_{\psi\psi\, (29)} &=& {\Gamma_{(1)}^q}_\alpha^{~\beta} {\Gamma_{(1)}^q}_\delta^{~\gamma} \delta^{ij} \delta^{kl} \delta_{IJ} \delta_{KL} 
~~~,~~~ 
{\cal P}_{\psi\psi\, (30)} ~=~ {\Gamma_{(1)}^q}_\alpha^{~\beta} {\Gamma_{(1)}^q}_\delta^{~\gamma} \delta^{ij} \delta^{kl} \delta_{IK} \delta_{JL} 
\nonumber \\
{\cal P}_{\psi\psi\, (31)} &=& {\Gamma_{(1)}^q}_\alpha^{~\beta} {\Gamma_{(1)}^r}_\delta^{~\gamma} \delta^{ij} \delta^{kl} \delta_{IJ} \delta_{KL} 
~~~,~~~ 
{\cal P}_{\psi\psi\, (32)} ~=~ {\Gamma_{(1)}^q}_\alpha^{~\beta} {\Gamma_{(1)}^r}_\delta^{~\gamma} \delta^{ij} \delta^{kl} \delta_{IK} \delta_{JL} 
\nonumber \\
{\cal P}_{\psi\psi\, (33)} &=& {\Gamma_{(1)}^q}_\alpha^{~\beta} {\Gamma_{(3)}^{p q r}}_\delta^{~\gamma} \delta^{ij} \delta^{kl} \delta_{IJ} \delta_{KL}
~~~,~~~ 
{\cal P}_{\psi\psi\, (34)} ~=~ {\Gamma_{(1)}^q}_\alpha^{~\beta} {\Gamma_{(3)}^{p q r}}_\delta^{~\gamma} \delta^{ij} \delta^{kl} \delta_{IK} \delta_{JL}
\nonumber \\
{\cal P}_{\psi\psi\, (35)} &=& {\Gamma_{(1)}^q}_\alpha^{~\gamma} {\Gamma_{(1)}^q}_\delta^{~\beta} \delta^{ik} \delta^{jl} \delta_{IJ} \delta_{KL} 
~~~,~~~ 
{\cal P}_{\psi\psi\, (36)} ~=~ {\Gamma_{(1)}^q}_\alpha^{~\gamma} {\Gamma_{(1)}^q}_\delta^{~\beta} \delta^{ik} \delta^{jl} \delta_{IK} \delta_{JL} 
\nonumber \\
{\cal P}_{\psi\psi\, (37)} &=& {\Gamma_{(1)}^q}_\alpha^{~\gamma} {\Gamma_{(1)}^r}_\delta^{~\beta} \delta^{ik} \delta^{jl} \delta_{IJ} \delta_{KL} 
~~~,~~~ 
{\cal P}_{\psi\psi\, (38)} ~=~ {\Gamma_{(1)}^q}_\alpha^{~\gamma} {\Gamma_{(1)}^r}_\delta^{~\beta} \delta^{ik} \delta^{jl} \delta_{IK} \delta_{JL} 
\nonumber \\
{\cal P}_{\psi\psi\, (39)} &=& {\Gamma_{(1)}^q}_\alpha^{~\gamma} {\Gamma_{(3)}^{p q r}}_\delta^{~\beta} \delta^{ik} \delta^{jl} \delta_{IJ} \delta_{KL}
~~~,~~~ 
{\cal P}_{\psi\psi\, (40)} ~=~ {\Gamma_{(1)}^q}_\alpha^{~\gamma} {\Gamma_{(3)}^{p q r}}_\delta^{~\beta} \delta^{ik} \delta^{jl} \delta_{IK} \delta_{JL}
\nonumber \\
{\cal P}_{\psi\psi\, (41)} &=& {\Gamma_{(1)}^q}_\delta^{~\beta} {\Gamma_{(1)}^r}_\alpha^{~\gamma} \delta^{ik} \delta^{jl} \delta_{IJ} \delta_{KL} 
~~~,~~~ 
{\cal P}_{\psi\psi\, (42)} ~=~ {\Gamma_{(1)}^q}_\delta^{~\beta} {\Gamma_{(1)}^r}_\alpha^{~\gamma} \delta^{ik} \delta^{jl} \delta_{IK} \delta_{JL} 
\nonumber \\
{\cal P}_{\psi\psi\, (43)} &=& {\Gamma_{(1)}^q}_\delta^{~\beta} {\Gamma_{(3)}^{p q r}}_\alpha^{~\gamma} \delta^{ik} \delta^{jl} \delta_{IJ} \delta_{KL}
~~~,~~~ 
{\cal P}_{\psi\psi\, (44)} ~=~ {\Gamma_{(1)}^q}_\delta^{~\beta} {\Gamma_{(3)}^{p q r}}_\alpha^{~\gamma} \delta^{ik} \delta^{jl} \delta_{IK} \delta_{JL}
\nonumber \\
{\cal P}_{\psi\psi\, (45)} &=& {\Gamma_{(1)}^q}_\delta^{~\gamma} {\Gamma_{(1)}^r}_\alpha^{~\beta} \delta^{ij} \delta^{kl} \delta_{IJ} \delta_{KL} 
~~~,~~~ 
{\cal P}_{\psi\psi\, (46)} ~=~ {\Gamma_{(1)}^q}_\delta^{~\gamma} {\Gamma_{(1)}^r}_\alpha^{~\beta} \delta^{ij} \delta^{kl} \delta_{IK} \delta_{JL} 
\nonumber \\
{\cal P}_{\psi\psi\, (47)} &=& {\Gamma_{(1)}^q}_\delta^{~\gamma} {\Gamma_{(3)}^{p q r}}_\alpha^{~\beta} \delta^{ij} \delta^{kl} \delta_{IJ} \delta_{KL}
~~~,~~~ 
{\cal P}_{\psi\psi\, (48)} ~=~ {\Gamma_{(1)}^q}_\delta^{~\gamma} {\Gamma_{(3)}^{p q r}}_\alpha^{~\beta} \delta^{ij} \delta^{kl} \delta_{IK} \delta_{JL}
\nonumber \\
{\cal P}_{\psi\psi\, (49)} &=& {\Gamma_{(1)}^r}_\alpha^{~\beta} {\Gamma_{(1)}^r}_\delta^{~\gamma} \delta^{ij} \delta^{kl} \delta_{IJ} \delta_{KL} 
~~~,~~~ 
{\cal P}_{\psi\psi\, (50)} ~=~ {\Gamma_{(1)}^r}_\alpha^{~\beta} {\Gamma_{(1)}^r}_\delta^{~\gamma} \delta^{ij} \delta^{kl} \delta_{IK} \delta_{JL} 
\nonumber \\
{\cal P}_{\psi\psi\, (51)} &=& {\Gamma_{(1)}^r}_\alpha^{~\beta} {\Gamma_{(3)}^{p q r}}_\delta^{~\gamma} \delta^{ij} \delta^{kl} \delta_{IJ} \delta_{KL}
~~~,~~~ 
{\cal P}_{\psi\psi\, (52)} ~=~ {\Gamma_{(1)}^r}_\alpha^{~\beta} {\Gamma_{(3)}^{p q r}}_\delta^{~\gamma} \delta^{ij} \delta^{kl} \delta_{IK} \delta_{JL}
\nonumber \\
{\cal P}_{\psi\psi\, (53)} &=& {\Gamma_{(1)}^r}_\alpha^{~\gamma} {\Gamma_{(1)}^r}_\delta^{~\beta} \delta^{ik} \delta^{jl} \delta_{IJ} \delta_{KL} 
~~~,~~~ 
{\cal P}_{\psi\psi\, (54)} ~=~ {\Gamma_{(1)}^r}_\alpha^{~\gamma} {\Gamma_{(1)}^r}_\delta^{~\beta} \delta^{ik} \delta^{jl} \delta_{IK} \delta_{JL} 
\nonumber \\
{\cal P}_{\psi\psi\, (55)} &=& {\Gamma_{(1)}^r}_\alpha^{~\gamma} {\Gamma_{(3)}^{p q r}}_\delta^{~\beta} \delta^{ik} \delta^{jl} \delta_{IJ} \delta_{KL}
~~~,~~~ 
{\cal P}_{\psi\psi\, (56)} ~=~ {\Gamma_{(1)}^r}_\alpha^{~\gamma} {\Gamma_{(3)}^{p q r}}_\delta^{~\beta} \delta^{ik} \delta^{jl} \delta_{IK} \delta_{JL}
\nonumber \\
{\cal P}_{\psi\psi\, (57)} &=& {\Gamma_{(1)}^r}_\delta^{~\beta} {\Gamma_{(3)}^{p q r}}_\alpha^{~\gamma} \delta^{ik} \delta^{jl} \delta_{IJ} \delta_{KL}
~~~,~~~ 
{\cal P}_{\psi\psi\, (58)} ~=~ {\Gamma_{(1)}^r}_\delta^{~\beta} {\Gamma_{(3)}^{p q r}}_\alpha^{~\gamma} \delta^{ik} \delta^{jl} \delta_{IK} \delta_{JL}
\nonumber \\
{\cal P}_{\psi\psi\, (59)} &=& {\Gamma_{(1)}^r}_\delta^{~\gamma} {\Gamma_{(3)}^{p q r}}_\alpha^{~\beta} \delta^{ij} \delta^{kl} \delta_{IJ} \delta_{KL}
~~~,~~~ 
{\cal P}_{\psi\psi\, (60)} ~=~ {\Gamma_{(1)}^r}_\delta^{~\gamma} {\Gamma_{(3)}^{p q r}}_\alpha^{~\beta} \delta^{ij} \delta^{kl} \delta_{IK} \delta_{JL}
\nonumber \\
{\cal P}_{\psi\psi\, (61)} &=& {\Gamma_{(1)}^\mu}_\alpha^{~\beta} {\Gamma_{(1)}^\mu}_\delta^{~\gamma} \delta^{ij} \delta^{kl} \delta_{IJ} \delta_{KL} 
~~~,~~~ 
{\cal P}_{\psi\psi\, (62)} ~=~ {\Gamma_{(1)}^\mu}_\alpha^{~\beta} {\Gamma_{(1)}^\mu}_\delta^{~\gamma} \delta^{ij} \delta^{kl} \delta_{IK} \delta_{JL} 
\nonumber \\
{\cal P}_{\psi\psi\, (63)} &=& {\Gamma_{(1)}^\mu}_\alpha^{~\beta} {\Gamma_{(3)}^{p q \mu}}_\delta^{~\gamma} \delta^{ij} \delta^{kl} \delta_{IJ} \delta_{KL}
~~~,~~~ 
{\cal P}_{\psi\psi\, (64)} ~=~ {\Gamma_{(1)}^\mu}_\alpha^{~\beta} {\Gamma_{(3)}^{p q \mu}}_\delta^{~\gamma} \delta^{ij} \delta^{kl} \delta_{IK} \delta_{JL}
\nonumber \\
{\cal P}_{\psi\psi\, (65)} &=& {\Gamma_{(1)}^\mu}_\alpha^{~\beta} {\Gamma_{(3)}^{p r \mu}}_\delta^{~\gamma} \delta^{ij} \delta^{kl} \delta_{IJ} \delta_{KL}
~~~,~~~ 
{\cal P}_{\psi\psi\, (66)} ~=~ {\Gamma_{(1)}^\mu}_\alpha^{~\beta} {\Gamma_{(3)}^{p r \mu}}_\delta^{~\gamma} \delta^{ij} \delta^{kl} \delta_{IK} \delta_{JL}
\nonumber \\
{\cal P}_{\psi\psi\, (67)} &=& {\Gamma_{(1)}^\mu}_\alpha^{~\beta} {\Gamma_{(3)}^{q r \mu}}_\delta^{~\gamma} \delta^{ij} \delta^{kl} \delta_{IJ} \delta_{KL}
~~~,~~~ 
{\cal P}_{\psi\psi\, (68)} ~=~ {\Gamma_{(1)}^\mu}_\alpha^{~\beta} {\Gamma_{(3)}^{q r \mu}}_\delta^{~\gamma} \delta^{ij} \delta^{kl} \delta_{IK} \delta_{JL}
\nonumber \\
{\cal P}_{\psi\psi\, (69)} &=& {\Gamma_{(1)}^\mu}_\alpha^{~\gamma} {\Gamma_{(1)}^\mu}_\delta^{~\beta} \delta^{ik} \delta^{jl} \delta_{IJ} \delta_{KL} 
~~~,~~~ 
{\cal P}_{\psi\psi\, (70)} ~=~ {\Gamma_{(1)}^\mu}_\alpha^{~\gamma} {\Gamma_{(1)}^\mu}_\delta^{~\beta} \delta^{ik} \delta^{jl} \delta_{IK} \delta_{JL} 
\nonumber \\
{\cal P}_{\psi\psi\, (71)} &=& {\Gamma_{(1)}^\mu}_\alpha^{~\gamma} {\Gamma_{(3)}^{p q \mu}}_\delta^{~\beta} \delta^{ik} \delta^{jl} \delta_{IJ} \delta_{KL}
~~~,~~~ 
{\cal P}_{\psi\psi\, (72)} ~=~ {\Gamma_{(1)}^\mu}_\alpha^{~\gamma} {\Gamma_{(3)}^{p q \mu}}_\delta^{~\beta} \delta^{ik} \delta^{jl} \delta_{IK} \delta_{JL}
\nonumber \\
{\cal P}_{\psi\psi\, (73)} &=& {\Gamma_{(1)}^\mu}_\alpha^{~\gamma} {\Gamma_{(3)}^{p r \mu}}_\delta^{~\beta} \delta^{ik} \delta^{jl} \delta_{IJ} \delta_{KL}
~~~,~~~ 
{\cal P}_{\psi\psi\, (74)} ~=~ {\Gamma_{(1)}^\mu}_\alpha^{~\gamma} {\Gamma_{(3)}^{p r \mu}}_\delta^{~\beta} \delta^{ik} \delta^{jl} \delta_{IK} \delta_{JL}
\nonumber \\
{\cal P}_{\psi\psi\, (75)} &=& {\Gamma_{(1)}^\mu}_\alpha^{~\gamma} {\Gamma_{(3)}^{q r \mu}}_\delta^{~\beta} \delta^{ik} \delta^{jl} \delta_{IJ} \delta_{KL}
~~~,~~~ 
{\cal P}_{\psi\psi\, (76)} ~=~ {\Gamma_{(1)}^\mu}_\alpha^{~\gamma} {\Gamma_{(3)}^{q r \mu}}_\delta^{~\beta} \delta^{ik} \delta^{jl} \delta_{IK} \delta_{JL}
\nonumber \\
{\cal P}_{\psi\psi\, (77)} &=& {\Gamma_{(1)}^\mu}_\delta^{~\beta} {\Gamma_{(3)}^{p q \mu}}_\alpha^{~\gamma} \delta^{ik} \delta^{jl} \delta_{IJ} \delta_{KL}
~~~,~~~ 
{\cal P}_{\psi\psi\, (78)} ~=~ {\Gamma_{(1)}^\mu}_\delta^{~\beta} {\Gamma_{(3)}^{p q \mu}}_\alpha^{~\gamma} \delta^{ik} \delta^{jl} \delta_{IK} \delta_{JL}
\nonumber \\
{\cal P}_{\psi\psi\, (79)} &=& {\Gamma_{(1)}^\mu}_\delta^{~\beta} {\Gamma_{(3)}^{p r \mu}}_\alpha^{~\gamma} \delta^{ik} \delta^{jl} \delta_{IJ} \delta_{KL}
~~~,~~~ 
{\cal P}_{\psi\psi\, (80)} ~=~ {\Gamma_{(1)}^\mu}_\delta^{~\beta} {\Gamma_{(3)}^{p r \mu}}_\alpha^{~\gamma} \delta^{ik} \delta^{jl} \delta_{IK} \delta_{JL}
\nonumber \\
{\cal P}_{\psi\psi\, (81)} &=& {\Gamma_{(1)}^\mu}_\delta^{~\beta} {\Gamma_{(3)}^{q r \mu}}_\alpha^{~\gamma} \delta^{ik} \delta^{jl} \delta_{IJ} \delta_{KL}
~~~,~~~ 
{\cal P}_{\psi\psi\, (82)} ~=~ {\Gamma_{(1)}^\mu}_\delta^{~\beta} {\Gamma_{(3)}^{q r \mu}}_\alpha^{~\gamma} \delta^{ik} \delta^{jl} \delta_{IK} \delta_{JL}
\nonumber \\
{\cal P}_{\psi\psi\, (83)} &=& {\Gamma_{(1)}^\mu}_\delta^{~\gamma} {\Gamma_{(3)}^{p q \mu}}_\alpha^{~\beta} \delta^{ij} \delta^{kl} \delta_{IJ} \delta_{KL}
~~~,~~~ 
{\cal P}_{\psi\psi\, (84)} ~=~ {\Gamma_{(1)}^\mu}_\delta^{~\gamma} {\Gamma_{(3)}^{p q \mu}}_\alpha^{~\beta} \delta^{ij} \delta^{kl} \delta_{IK} \delta_{JL}
\nonumber \\
{\cal P}_{\psi\psi\, (85)} &=& {\Gamma_{(1)}^\mu}_\delta^{~\gamma} {\Gamma_{(3)}^{p r \mu}}_\alpha^{~\beta} \delta^{ij} \delta^{kl} \delta_{IJ} \delta_{KL}
~~~,~~~ 
{\cal P}_{\psi\psi\, (86)} ~=~ {\Gamma_{(1)}^\mu}_\delta^{~\gamma} {\Gamma_{(3)}^{p r \mu}}_\alpha^{~\beta} \delta^{ij} \delta^{kl} \delta_{IK} \delta_{JL}
\nonumber \\
{\cal P}_{\psi\psi\, (87)} &=& {\Gamma_{(1)}^\mu}_\delta^{~\gamma} {\Gamma_{(3)}^{q r \mu}}_\alpha^{~\beta} \delta^{ij} \delta^{kl} \delta_{IJ} \delta_{KL}
~~~,~~~ 
{\cal P}_{\psi\psi\, (88)} ~=~ {\Gamma_{(1)}^\mu}_\delta^{~\gamma} {\Gamma_{(3)}^{q r \mu}}_\alpha^{~\beta} \delta^{ij} \delta^{kl} \delta_{IK} \delta_{JL}
\nonumber \\
{\cal P}_{\psi\psi\, (89)} &=& {\Gamma_{(3)}^{p q r}}_\alpha^{~\beta} {\Gamma_{(3)}^{p q r}}_\delta^{~\gamma} \delta^{ij} \delta^{kl} \delta_{IJ} \delta_{KL}
~~~,~~~ 
{\cal P}_{\psi\psi\, (90)} ~=~ {\Gamma_{(3)}^{p q r}}_\alpha^{~\beta} {\Gamma_{(3)}^{p q r}}_\delta^{~\gamma} \delta^{ij} \delta^{kl} \delta_{IK} \delta_{JL}
\nonumber \\
{\cal P}_{\psi\psi\, (91)} &=& {\Gamma_{(3)}^{p q r}}_\alpha^{~\gamma} {\Gamma_{(3)}^{p q r}}_\delta^{~\beta} \delta^{ik} \delta^{jl} \delta_{IJ} \delta_{KL}
~~~,~~~ 
{\cal P}_{\psi\psi\, (92)} ~=~ {\Gamma_{(3)}^{p q r}}_\alpha^{~\gamma} {\Gamma_{(3)}^{p q r}}_\delta^{~\beta} \delta^{ik} \delta^{jl} \delta_{IK} \delta_{JL}
\nonumber \\
{\cal P}_{\psi\psi\, (93)} &=& {\Gamma_{(3)}^{p q \sigma}}_\alpha^{~\beta} {\Gamma_{(3)}^{p q \sigma}}_\delta^{~\gamma} \delta^{ij} \delta^{kl} \delta_{IJ} \delta_{KL}
~~~,~~~ 
{\cal P}_{\psi\psi\, (94)} ~=~ {\Gamma_{(3)}^{p q \sigma}}_\alpha^{~\beta} {\Gamma_{(3)}^{p q \sigma}}_\delta^{~\gamma} \delta^{ij} \delta^{kl} \delta_{IK} \delta_{JL}
\nonumber \\
{\cal P}_{\psi\psi\, (95)} &=& {\Gamma_{(3)}^{p q \sigma}}_\alpha^{~\beta} {\Gamma_{(3)}^{p r \sigma}}_\delta^{~\gamma} \delta^{ij} \delta^{kl} \delta_{IJ} \delta_{KL}
~~~,~~~ 
{\cal P}_{\psi\psi\, (96)} ~=~ {\Gamma_{(3)}^{p q \sigma}}_\alpha^{~\beta} {\Gamma_{(3)}^{p r \sigma}}_\delta^{~\gamma} \delta^{ij} \delta^{kl} \delta_{IK} \delta_{JL}
\nonumber \\
{\cal P}_{\psi\psi\, (97)} &=& {\Gamma_{(3)}^{p q \sigma}}_\alpha^{~\beta} {\Gamma_{(3)}^{q r \sigma}}_\delta^{~\gamma} \delta^{ij} \delta^{kl} \delta_{IJ} \delta_{KL}
~~~,~~~ 
{\cal P}_{\psi\psi\, (98)} ~=~ {\Gamma_{(3)}^{p q \sigma}}_\alpha^{~\beta} {\Gamma_{(3)}^{q r \sigma}}_\delta^{~\gamma} \delta^{ij} \delta^{kl} \delta_{IK} \delta_{JL}
\nonumber \\
{\cal P}_{\psi\psi\, (99)} &=& {\Gamma_{(3)}^{p q \sigma}}_\alpha^{~\gamma} {\Gamma_{(3)}^{p q \sigma}}_\delta^{~\beta} \delta^{ik} \delta^{jl} \delta_{IJ} \delta_{KL}
~~~,~~~ 
{\cal P}_{\psi\psi\, (100)} ~=~ {\Gamma_{(3)}^{p q \sigma}}_\alpha^{~\gamma} {\Gamma_{(3)}^{p q \sigma}}_\delta^{~\beta} \delta^{ik} \delta^{jl} \delta_{IK} \delta_{JL}
\nonumber \\
{\cal P}_{\psi\psi\, (101)} &=& {\Gamma_{(3)}^{p q \sigma}}_\alpha^{~\gamma} {\Gamma_{(3)}^{p r \sigma}}_\delta^{~\beta} \delta^{ik} \delta^{jl} \delta_{IJ} \delta_{KL}
~~~,~~~ 
{\cal P}_{\psi\psi\, (102)} ~=~ {\Gamma_{(3)}^{p q \sigma}}_\alpha^{~\gamma} {\Gamma_{(3)}^{p r \sigma}}_\delta^{~\beta} \delta^{ik} \delta^{jl} \delta_{IK} \delta_{JL}
\nonumber \\
{\cal P}_{\psi\psi\, (103)} &=& {\Gamma_{(3)}^{p q \sigma}}_\alpha^{~\gamma} {\Gamma_{(3)}^{q r \sigma}}_\delta^{~\beta} \delta^{ik} \delta^{jl} \delta_{IJ} \delta_{KL}
~~~,~~~ 
{\cal P}_{\psi\psi\, (104)} ~=~ {\Gamma_{(3)}^{p q \sigma}}_\alpha^{~\gamma} {\Gamma_{(3)}^{q r \sigma}}_\delta^{~\beta} \delta^{ik} \delta^{jl} \delta_{IK} \delta_{JL}
\nonumber \\
{\cal P}_{\psi\psi\, (105)} &=& {\Gamma_{(3)}^{p q \sigma}}_\delta^{~\beta} {\Gamma_{(3)}^{p r \sigma}}_\alpha^{~\gamma} \delta^{ik} \delta^{jl} \delta_{IJ} \delta_{KL}
~~~,~~~ 
{\cal P}_{\psi\psi\, (106)} ~=~ {\Gamma_{(3)}^{p q \sigma}}_\delta^{~\beta} {\Gamma_{(3)}^{p r \sigma}}_\alpha^{~\gamma} \delta^{ik} \delta^{jl} \delta_{IK} \delta_{JL}
\nonumber \\
{\cal P}_{\psi\psi\, (107)} &=& {\Gamma_{(3)}^{p q \sigma}}_\delta^{~\beta} {\Gamma_{(3)}^{q r \sigma}}_\alpha^{~\gamma} \delta^{ik} \delta^{jl} \delta_{IJ} \delta_{KL}
~~~,~~~ 
{\cal P}_{\psi\psi\, (108)} ~=~ {\Gamma_{(3)}^{p q \sigma}}_\delta^{~\beta} {\Gamma_{(3)}^{q r \sigma}}_\alpha^{~\gamma} \delta^{ik} \delta^{jl} \delta_{IK} \delta_{JL}
\nonumber \\
{\cal P}_{\psi\psi\, (109)} &=& {\Gamma_{(3)}^{p q \sigma}}_\delta^{~\gamma} {\Gamma_{(3)}^{p r \sigma}}_\alpha^{~\beta} \delta^{ij} \delta^{kl} \delta_{IJ} \delta_{KL}
~~~,~~~ 
{\cal P}_{\psi\psi\, (110)} ~=~ {\Gamma_{(3)}^{p q \sigma}}_\delta^{~\gamma} {\Gamma_{(3)}^{p r \sigma}}_\alpha^{~\beta} \delta^{ij} \delta^{kl} \delta_{IK} \delta_{JL}
\nonumber \\
{\cal P}_{\psi\psi\, (111)} &=& {\Gamma_{(3)}^{p q \sigma}}_\delta^{~\gamma} {\Gamma_{(3)}^{q r \sigma}}_\alpha^{~\beta} \delta^{ij} \delta^{kl} \delta_{IJ} \delta_{KL}
~~~,~~~ 
{\cal P}_{\psi\psi\, (112)} ~=~ {\Gamma_{(3)}^{p q \sigma}}_\delta^{~\gamma} {\Gamma_{(3)}^{q r \sigma}}_\alpha^{~\beta} \delta^{ij} \delta^{kl} \delta_{IK} \delta_{JL}
\nonumber \\
{\cal P}_{\psi\psi\, (113)} &=& {\Gamma_{(3)}^{p r \sigma}}_\alpha^{~\beta} {\Gamma_{(3)}^{p r \sigma}}_\delta^{~\gamma} \delta^{ij} \delta^{kl} \delta_{IJ} \delta_{KL}
~~~,~~~ 
{\cal P}_{\psi\psi\, (114)} ~=~ {\Gamma_{(3)}^{p r \sigma}}_\alpha^{~\beta} {\Gamma_{(3)}^{p r \sigma}}_\delta^{~\gamma} \delta^{ij} \delta^{kl} \delta_{IK} \delta_{JL}
\nonumber \\
{\cal P}_{\psi\psi\, (115)} &=& {\Gamma_{(3)}^{p r \sigma}}_\alpha^{~\beta} {\Gamma_{(3)}^{q r \sigma}}_\delta^{~\gamma} \delta^{ij} \delta^{kl} \delta_{IJ} \delta_{KL}
~~~,~~~ 
{\cal P}_{\psi\psi\, (116)} ~=~ {\Gamma_{(3)}^{p r \sigma}}_\alpha^{~\beta} {\Gamma_{(3)}^{q r \sigma}}_\delta^{~\gamma} \delta^{ij} \delta^{kl} \delta_{IK} \delta_{JL}
\nonumber \\
{\cal P}_{\psi\psi\, (117)} &=& {\Gamma_{(3)}^{p r \sigma}}_\alpha^{~\gamma} {\Gamma_{(3)}^{p r \sigma}}_\delta^{~\beta} \delta^{ik} \delta^{jl} \delta_{IJ} \delta_{KL}
~~~,~~~ 
{\cal P}_{\psi\psi\, (118)} ~=~ {\Gamma_{(3)}^{p r \sigma}}_\alpha^{~\gamma} {\Gamma_{(3)}^{p r \sigma}}_\delta^{~\beta} \delta^{ik} \delta^{jl} \delta_{IK} \delta_{JL}
\nonumber \\
{\cal P}_{\psi\psi\, (119)} &=& {\Gamma_{(3)}^{p r \sigma}}_\alpha^{~\gamma} {\Gamma_{(3)}^{q r \sigma}}_\delta^{~\beta} \delta^{ik} \delta^{jl} \delta_{IJ} \delta_{KL}
~~~,~~~ 
{\cal P}_{\psi\psi\, (120)} ~=~ {\Gamma_{(3)}^{p r \sigma}}_\alpha^{~\gamma} {\Gamma_{(3)}^{q r \sigma}}_\delta^{~\beta} \delta^{ik} \delta^{jl} \delta_{IK} \delta_{JL}
\nonumber \\
{\cal P}_{\psi\psi\, (121)} &=& {\Gamma_{(3)}^{p r \sigma}}_\delta^{~\beta} {\Gamma_{(3)}^{q r \sigma}}_\alpha^{~\gamma} \delta^{ik} \delta^{jl} \delta_{IJ} \delta_{KL}
~~~,~~~ 
{\cal P}_{\psi\psi\, (122)} ~=~ {\Gamma_{(3)}^{p r \sigma}}_\delta^{~\beta} {\Gamma_{(3)}^{q r \sigma}}_\alpha^{~\gamma} \delta^{ik} \delta^{jl} \delta_{IK} \delta_{JL}
\nonumber \\
{\cal P}_{\psi\psi\, (123)} &=& {\Gamma_{(3)}^{p r \sigma}}_\delta^{~\gamma} {\Gamma_{(3)}^{q r \sigma}}_\alpha^{~\beta} \delta^{ij} \delta^{kl} \delta_{IJ} \delta_{KL}
~~~,~~~ 
{\cal P}_{\psi\psi\, (124)} ~=~ {\Gamma_{(3)}^{p r \sigma}}_\delta^{~\gamma} {\Gamma_{(3)}^{q r \sigma}}_\alpha^{~\beta} \delta^{ij} \delta^{kl} \delta_{IK} \delta_{JL}
\nonumber \\
{\cal P}_{\psi\psi\, (125)} &=& {\Gamma_{(3)}^{p \nu \sigma}}_\alpha^{~\beta} {\Gamma_{(3)}^{p \nu \sigma}}_\delta^{~\gamma} \delta^{ij} \delta^{kl} \delta_{IJ} \delta_{KL}
~~~,~~~ 
{\cal P}_{\psi\psi\, (126)} ~=~ {\Gamma_{(3)}^{p \nu \sigma}}_\alpha^{~\beta} {\Gamma_{(3)}^{p \nu \sigma}}_\delta^{~\gamma} \delta^{ij} \delta^{kl} \delta_{IK} \delta_{JL}
\nonumber \\
{\cal P}_{\psi\psi\, (127)} &=& {\Gamma_{(3)}^{p \nu \sigma}}_\alpha^{~\beta} {\Gamma_{(3)}^{q \nu \sigma}}_\delta^{~\gamma} \delta^{ij} \delta^{kl} \delta_{IJ} \delta_{KL}
~~~,~~~ 
{\cal P}_{\psi\psi\, (128)} ~=~ {\Gamma_{(3)}^{p \nu \sigma}}_\alpha^{~\beta} {\Gamma_{(3)}^{q \nu \sigma}}_\delta^{~\gamma} \delta^{ij} \delta^{kl} \delta_{IK} \delta_{JL}
\nonumber \\
{\cal P}_{\psi\psi\, (129)} &=& {\Gamma_{(3)}^{p \nu \sigma}}_\alpha^{~\beta} {\Gamma_{(3)}^{r \nu \sigma}}_\delta^{~\gamma} \delta^{ij} \delta^{kl} \delta_{IJ} \delta_{KL}
~~~,~~~ 
{\cal P}_{\psi\psi\, (130)} ~=~ {\Gamma_{(3)}^{p \nu \sigma}}_\alpha^{~\beta} {\Gamma_{(3)}^{r \nu \sigma}}_\delta^{~\gamma} \delta^{ij} \delta^{kl} \delta_{IK} \delta_{JL}
\nonumber \\
{\cal P}_{\psi\psi\, (131)} &=& {\Gamma_{(3)}^{p \nu \sigma}}_\alpha^{~\gamma} {\Gamma_{(3)}^{p \nu \sigma}}_\delta^{~\beta} \delta^{ik} \delta^{jl} \delta_{IJ} \delta_{KL}
~~~,~~~ 
{\cal P}_{\psi\psi\, (132)} ~=~ {\Gamma_{(3)}^{p \nu \sigma}}_\alpha^{~\gamma} {\Gamma_{(3)}^{p \nu \sigma}}_\delta^{~\beta} \delta^{ik} \delta^{jl} \delta_{IK} \delta_{JL}
\nonumber \\
{\cal P}_{\psi\psi\, (133)} &=& {\Gamma_{(3)}^{p \nu \sigma}}_\alpha^{~\gamma} {\Gamma_{(3)}^{q \nu \sigma}}_\delta^{~\beta} \delta^{ik} \delta^{jl} \delta_{IJ} \delta_{KL}
~~~,~~~ 
{\cal P}_{\psi\psi\, (134)} ~=~ {\Gamma_{(3)}^{p \nu \sigma}}_\alpha^{~\gamma} {\Gamma_{(3)}^{q \nu \sigma}}_\delta^{~\beta} \delta^{ik} \delta^{jl} \delta_{IK} \delta_{JL}
\nonumber \\
{\cal P}_{\psi\psi\, (135)} &=& {\Gamma_{(3)}^{p \nu \sigma}}_\alpha^{~\gamma} {\Gamma_{(3)}^{r \nu \sigma}}_\delta^{~\beta} \delta^{ik} \delta^{jl} \delta_{IJ} \delta_{KL}
~~~,~~~ 
{\cal P}_{\psi\psi\, (136)} ~=~ {\Gamma_{(3)}^{p \nu \sigma}}_\alpha^{~\gamma} {\Gamma_{(3)}^{r \nu \sigma}}_\delta^{~\beta} \delta^{ik} \delta^{jl} \delta_{IK} \delta_{JL}
\nonumber \\
{\cal P}_{\psi\psi\, (137)} &=& {\Gamma_{(3)}^{p \nu \sigma}}_\delta^{~\beta} {\Gamma_{(3)}^{q \nu \sigma}}_\alpha^{~\gamma} \delta^{ik} \delta^{jl} \delta_{IJ} \delta_{KL}
~~~,~~~ 
{\cal P}_{\psi\psi\, (138)} ~=~ {\Gamma_{(3)}^{p \nu \sigma}}_\delta^{~\beta} {\Gamma_{(3)}^{q \nu \sigma}}_\alpha^{~\gamma} \delta^{ik} \delta^{jl} \delta_{IK} \delta_{JL}
\nonumber \\
{\cal P}_{\psi\psi\, (139)} &=& {\Gamma_{(3)}^{p \nu \sigma}}_\delta^{~\beta} {\Gamma_{(3)}^{r \nu \sigma}}_\alpha^{~\gamma} \delta^{ik} \delta^{jl} \delta_{IJ} \delta_{KL}
~~~,~~~ 
{\cal P}_{\psi\psi\, (140)} ~=~ {\Gamma_{(3)}^{p \nu \sigma}}_\delta^{~\beta} {\Gamma_{(3)}^{r \nu \sigma}}_\alpha^{~\gamma} \delta^{ik} \delta^{jl} \delta_{IK} \delta_{JL}
\nonumber \\
{\cal P}_{\psi\psi\, (141)} &=& {\Gamma_{(3)}^{p \nu \sigma}}_\delta^{~\gamma} {\Gamma_{(3)}^{q \nu \sigma}}_\alpha^{~\beta} \delta^{ij} \delta^{kl} \delta_{IJ} \delta_{KL}
~~~,~~~ 
{\cal P}_{\psi\psi\, (142)} ~=~ {\Gamma_{(3)}^{p \nu \sigma}}_\delta^{~\gamma} {\Gamma_{(3)}^{q \nu \sigma}}_\alpha^{~\beta} \delta^{ij} \delta^{kl} \delta_{IK} \delta_{JL}
\nonumber \\
{\cal P}_{\psi\psi\, (143)} &=& {\Gamma_{(3)}^{p \nu \sigma}}_\delta^{~\gamma} {\Gamma_{(3)}^{r \nu \sigma}}_\alpha^{~\beta} \delta^{ij} \delta^{kl} \delta_{IJ} \delta_{KL}
~~~,~~~ 
{\cal P}_{\psi\psi\, (144)} ~=~ {\Gamma_{(3)}^{p \nu \sigma}}_\delta^{~\gamma} {\Gamma_{(3)}^{r \nu \sigma}}_\alpha^{~\beta} \delta^{ij} \delta^{kl} \delta_{IK} \delta_{JL}
\nonumber \\
{\cal P}_{\psi\psi\, (145)} &=& {\Gamma_{(3)}^{q r \sigma}}_\alpha^{~\beta} {\Gamma_{(3)}^{q r \sigma}}_\delta^{~\gamma} \delta^{ij} \delta^{kl} \delta_{IJ} \delta_{KL}
~~~,~~~ 
{\cal P}_{\psi\psi\, (146)} ~=~ {\Gamma_{(3)}^{q r \sigma}}_\alpha^{~\beta} {\Gamma_{(3)}^{q r \sigma}}_\delta^{~\gamma} \delta^{ij} \delta^{kl} \delta_{IK} \delta_{JL}
\nonumber \\
{\cal P}_{\psi\psi\, (147)} &=& {\Gamma_{(3)}^{q r \sigma}}_\alpha^{~\gamma} {\Gamma_{(3)}^{q r \sigma}}_\delta^{~\beta} \delta^{ik} \delta^{jl} \delta_{IJ} \delta_{KL}
~~~,~~~ 
{\cal P}_{\psi\psi\, (148)} ~=~ {\Gamma_{(3)}^{q r \sigma}}_\alpha^{~\gamma} {\Gamma_{(3)}^{q r \sigma}}_\delta^{~\beta} \delta^{ik} \delta^{jl} \delta_{IK} \delta_{JL}
\nonumber \\
{\cal P}_{\psi\psi\, (149)} &=& {\Gamma_{(3)}^{q \nu \sigma}}_\alpha^{~\beta} {\Gamma_{(3)}^{q \nu \sigma}}_\delta^{~\gamma} \delta^{ij} \delta^{kl} \delta_{IJ} \delta_{KL}
~~~,~~~ 
{\cal P}_{\psi\psi\, (150)} ~=~ {\Gamma_{(3)}^{q \nu \sigma}}_\alpha^{~\beta} {\Gamma_{(3)}^{q \nu \sigma}}_\delta^{~\gamma} \delta^{ij} \delta^{kl} \delta_{IK} \delta_{JL}
\nonumber \\
{\cal P}_{\psi\psi\, (151)} &=& {\Gamma_{(3)}^{q \nu \sigma}}_\alpha^{~\beta} {\Gamma_{(3)}^{r \nu \sigma}}_\delta^{~\gamma} \delta^{ij} \delta^{kl} \delta_{IJ} \delta_{KL}
~~~,~~~ 
{\cal P}_{\psi\psi\, (152)} ~=~ {\Gamma_{(3)}^{q \nu \sigma}}_\alpha^{~\beta} {\Gamma_{(3)}^{r \nu \sigma}}_\delta^{~\gamma} \delta^{ij} \delta^{kl} \delta_{IK} \delta_{JL}
\nonumber \\
{\cal P}_{\psi\psi\, (153)} &=& {\Gamma_{(3)}^{q \nu \sigma}}_\alpha^{~\gamma} {\Gamma_{(3)}^{q \nu \sigma}}_\delta^{~\beta} \delta^{ik} \delta^{jl} \delta_{IJ} \delta_{KL}
~~~,~~~ 
{\cal P}_{\psi\psi\, (154)} ~=~ {\Gamma_{(3)}^{q \nu \sigma}}_\alpha^{~\gamma} {\Gamma_{(3)}^{q \nu \sigma}}_\delta^{~\beta} \delta^{ik} \delta^{jl} \delta_{IK} \delta_{JL}
\nonumber \\
{\cal P}_{\psi\psi\, (155)} &=& {\Gamma_{(3)}^{q \nu \sigma}}_\alpha^{~\gamma} {\Gamma_{(3)}^{r \nu \sigma}}_\delta^{~\beta} \delta^{ik} \delta^{jl} \delta_{IJ} \delta_{KL}
~~~,~~~ 
{\cal P}_{\psi\psi\, (156)} ~=~ {\Gamma_{(3)}^{q \nu \sigma}}_\alpha^{~\gamma} {\Gamma_{(3)}^{r \nu \sigma}}_\delta^{~\beta} \delta^{ik} \delta^{jl} \delta_{IK} \delta_{JL}
\nonumber \\
{\cal P}_{\psi\psi\, (157)} &=& {\Gamma_{(3)}^{q \nu \sigma}}_\delta^{~\beta} {\Gamma_{(3)}^{r \nu \sigma}}_\alpha^{~\gamma} \delta^{ik} \delta^{jl} \delta_{IJ} \delta_{KL}
~~~,~~~ 
{\cal P}_{\psi\psi\, (158)} ~=~ {\Gamma_{(3)}^{q \nu \sigma}}_\delta^{~\beta} {\Gamma_{(3)}^{r \nu \sigma}}_\alpha^{~\gamma} \delta^{ik} \delta^{jl} \delta_{IK} \delta_{JL}
\nonumber \\
{\cal P}_{\psi\psi\, (159)} &=& {\Gamma_{(3)}^{q \nu \sigma}}_\delta^{~\gamma} {\Gamma_{(3)}^{r \nu \sigma}}_\alpha^{~\beta} \delta^{ij} \delta^{kl} \delta_{IJ} \delta_{KL}
~~~,~~~ 
{\cal P}_{\psi\psi\, (160)} ~=~ {\Gamma_{(3)}^{q \nu \sigma}}_\delta^{~\gamma} {\Gamma_{(3)}^{r \nu \sigma}}_\alpha^{~\beta} \delta^{ij} \delta^{kl} \delta_{IK} \delta_{JL}
\nonumber \\
{\cal P}_{\psi\psi\, (161)} &=& {\Gamma_{(3)}^{r \nu \sigma}}_\alpha^{~\beta} {\Gamma_{(3)}^{r \nu \sigma}}_\delta^{~\gamma} \delta^{ij} \delta^{kl} \delta_{IJ} \delta_{KL}
~~~,~~~ 
{\cal P}_{\psi\psi\, (162)} ~=~ {\Gamma_{(3)}^{r \nu \sigma}}_\alpha^{~\beta} {\Gamma_{(3)}^{r \nu \sigma}}_\delta^{~\gamma} \delta^{ij} \delta^{kl} \delta_{IK} \delta_{JL}
\nonumber \\
{\cal P}_{\psi\psi\, (163)} &=& {\Gamma_{(3)}^{r \nu \sigma}}_\alpha^{~\gamma} {\Gamma_{(3)}^{r \nu \sigma}}_\delta^{~\beta} \delta^{ik} \delta^{jl} \delta_{IJ} \delta_{KL}
~~~,~~~ 
{\cal P}_{\psi\psi\, (164)} ~=~ {\Gamma_{(3)}^{r \nu \sigma}}_\alpha^{~\gamma} {\Gamma_{(3)}^{r \nu \sigma}}_\delta^{~\beta} \delta^{ik} \delta^{jl} \delta_{IK} \delta_{JL}
\nonumber \\
{\cal P}_{\psi\psi\, (165)} &=& {\Gamma_{(3)}^{\mu \nu \sigma}}_\alpha^{~\beta} {\Gamma_{(3)}^{\mu \nu \sigma}}_\delta^{~\gamma} \delta^{ij} \delta^{kl} \delta_{IJ} \delta_{KL}
~~~,~~~ 
{\cal P}_{\psi\psi\, (166)} ~=~ {\Gamma_{(3)}^{\mu \nu \sigma}}_\alpha^{~\beta} {\Gamma_{(3)}^{\mu \nu \sigma}}_\delta^{~\gamma} \delta^{ij} \delta^{kl} \delta_{IK} \delta_{JL}
\nonumber \\
{\cal P}_{\psi\psi\, (167)} &=& {\Gamma_{(3)}^{\mu \nu \sigma}}_\alpha^{~\gamma} {\Gamma_{(3)}^{\mu \nu \sigma}}_\delta^{~\beta} \delta^{ik} \delta^{jl} \delta_{IJ} \delta_{KL}
~~~,~~~ 
{\cal P}_{\psi\psi\, (168)} ~=~ {\Gamma_{(3)}^{\mu \nu \sigma}}_\alpha^{~\gamma} {\Gamma_{(3)}^{\mu \nu \sigma}}_\delta^{~\beta} \delta^{ik} \delta^{jl} \delta_{IK} \delta_{JL} ~~~~
\end{eqnarray}
where we have suppressed the mass scale $\mu$ for presentational reasons. It
can be restored by using a dimensional argument. When higher loop corrections 
are included in this Green's function we expect the basis to be larger. For 
instance tensors such as 
$\Gamma_{(5)}^{\mu\nu pqr}$~$\otimes$~$\Gamma_{(5)}^{\mu\nu pqr}$ should appear
at two loops given the number of $\gamma$-matrices which would arise. 

By contrast the basis for the quartic electron Green's function in QED is a
smaller basis than that for the quark. This is partly due to the absence of the
colour group theory but also due to a smaller number of contributing Feynman 
graphs. We have 
\begin{eqnarray}
{\cal P}^{\QEDs}_{\psi\psi\, (1)} &=& {\Gamma_{(1)}^{p}}_{\alpha}^{~\beta} {\Gamma_{(1)}^{p}}_{\delta}^{~\gamma} \delta^{ij} \delta^{kl} ~~~,~~~ 
{\cal P}^{\QEDs}_{\psi\psi\, (2)} ~=~ {\Gamma_{(1)}^{p}}_{\alpha}^{~\beta} {\Gamma_{(1)}^{q}}_{\delta}^{~\gamma} \delta^{ij} \delta^{kl} 
\nonumber \\
{\cal P}^{\QEDs}_{\psi\psi\, (3)} &=& {\Gamma_{(1)}^{p}}_{\alpha}^{~\beta} {\Gamma_{(1)}^{r}}_{\delta}^{~\gamma} \delta^{ij} \delta^{kl} ~~~,~~~ 
{\cal P}^{\QEDs}_{\psi\psi\, (4)} ~=~ {\Gamma_{(1)}^{p}}_{\alpha}^{~\beta} {\Gamma_{(3)}^{p q r}}_{\delta}^{~\gamma} \delta^{ij} \delta^{kl} 
\nonumber \\
{\cal P}^{\QEDs}_{\psi\psi\, (5)} &=& {\Gamma_{(1)}^{p}}_{\alpha}^{~\gamma} {\Gamma_{(1)}^{q}}_{\delta}^{~\beta} \delta^{ik} \delta^{jl} ~~~,~~~ 
{\cal P}^{\QEDs}_{\psi\psi\, (6)} ~=~ {\Gamma_{(1)}^{p}}_{\alpha}^{~\gamma} {\Gamma_{(1)}^{r}}_{\delta}^{~\beta} \delta^{ik} \delta^{jl} 
\nonumber \\
{\cal P}^{\QEDs}_{\psi\psi\, (7)} &=& {\Gamma_{(1)}^{p}}_{\alpha}^{~\gamma} {\Gamma_{(3)}^{p q r}}_{\delta}^{~\beta} \delta^{ik} \delta^{jl} ~~~,~~~ 
{\cal P}^{\QEDs}_{\psi\psi\, (8)} ~=~ {\Gamma_{(1)}^{p}}_{\delta}^{~\beta} {\Gamma_{(1)}^{q}}_{\alpha}^{~\gamma} \delta^{ik} \delta^{jl} 
\nonumber \\
{\cal P}^{\QEDs}_{\psi\psi\, (9)} &=& {\Gamma_{(1)}^{p}}_{\delta}^{~\beta} {\Gamma_{(1)}^{r}}_{\alpha}^{~\gamma} \delta^{ik} \delta^{jl} ~~~,~~~ 
{\cal P}^{\QEDs}_{\psi\psi\, (10)} ~=~ {\Gamma_{(1)}^{p}}_{\delta}^{~\gamma} {\Gamma_{(1)}^{q}}_{\alpha}^{~\beta} \delta^{ij} \delta^{kl} 
\nonumber \\
{\cal P}^{\QEDs}_{\psi\psi\, (11)} &=& {\Gamma_{(1)}^{p}}_{\delta}^{~\gamma} {\Gamma_{(1)}^{r}}_{\alpha}^{~\beta} \delta^{ij} \delta^{kl} ~~~,~~~ 
{\cal P}^{\QEDs}_{\psi\psi\, (12)} ~=~ {\Gamma_{(1)}^{p}}_{\delta}^{~\gamma} {\Gamma_{(3)}^{p q r}}_{\alpha}^{~\beta} \delta^{ij} \delta^{kl} 
\nonumber \\
{\cal P}^{\QEDs}_{\psi\psi\, (13)} &=& {\Gamma_{(1)}^{q}}_{\alpha}^{~\beta} {\Gamma_{(1)}^{q}}_{\delta}^{~\gamma} \delta^{ij} \delta^{kl} ~~~,~~~ 
{\cal P}^{\QEDs}_{\psi\psi\, (14)} ~=~ {\Gamma_{(1)}^{q}}_{\alpha}^{~\beta} {\Gamma_{(1)}^{r}}_{\delta}^{~\gamma} \delta^{ij} \delta^{kl} 
\nonumber \\
{\cal P}^{\QEDs}_{\psi\psi\, (15)} &=& {\Gamma_{(1)}^{q}}_{\alpha}^{~\beta} {\Gamma_{(3)}^{p q r}}_{\delta}^{~\gamma} \delta^{ij} \delta^{kl} ~~~,~~~ 
{\cal P}^{\QEDs}_{\psi\psi\, (16)} ~=~ {\Gamma_{(1)}^{q}}_{\alpha}^{~\gamma} {\Gamma_{(1)}^{q}}_{\delta}^{~\beta} \delta^{ik} \delta^{jl} 
\nonumber \\
{\cal P}^{\QEDs}_{\psi\psi\, (17)} &=& {\Gamma_{(1)}^{q}}_{\alpha}^{~\gamma} {\Gamma_{(1)}^{r}}_{\delta}^{~\beta} \delta^{ik} \delta^{jl} ~~~,~~~ 
{\cal P}^{\QEDs}_{\psi\psi\, (18)} ~=~ {\Gamma_{(1)}^{q}}_{\alpha}^{~\gamma} {\Gamma_{(3)}^{p q r}}_{\delta}^{~\beta} \delta^{ik} \delta^{jl} 
\nonumber \\
{\cal P}^{\QEDs}_{\psi\psi\, (19)} &=& {\Gamma_{(1)}^{q}}_{\delta}^{~\beta} {\Gamma_{(1)}^{r}}_{\alpha}^{~\gamma} \delta^{ik} \delta^{jl} ~~~,~~~ 
{\cal P}^{\QEDs}_{\psi\psi\, (20)} ~=~ {\Gamma_{(1)}^{q}}_{\delta}^{~\beta} {\Gamma_{(3)}^{p q r}}_{\alpha}^{~\gamma} \delta^{ik} \delta^{jl} 
\nonumber \\
{\cal P}^{\QEDs}_{\psi\psi\, (21)} &=& {\Gamma_{(1)}^{q}}_{\delta}^{~\gamma} {\Gamma_{(1)}^{r}}_{\alpha}^{~\beta} \delta^{ij} \delta^{kl} ~~~,~~~ 
{\cal P}^{\QEDs}_{\psi\psi\, (22)} ~=~ {\Gamma_{(1)}^{q}}_{\delta}^{~\gamma} {\Gamma_{(3)}^{p q r}}_{\alpha}^{~\beta} \delta^{ij} \delta^{kl} 
\nonumber \\
{\cal P}^{\QEDs}_{\psi\psi\, (23)} &=& {\Gamma_{(1)}^{r}}_{\alpha}^{~\beta} {\Gamma_{(3)}^{p q r}}_{\delta}^{~\gamma} \delta^{ij} \delta^{kl} ~~~,~~~ 
{\cal P}^{\QEDs}_{\psi\psi\, (24)} ~=~ {\Gamma_{(1)}^{r}}_{\alpha}^{~\gamma} {\Gamma_{(1)}^{r}}_{\delta}^{~\beta} \delta^{ik} \delta^{jl} 
\nonumber \\
{\cal P}^{\QEDs}_{\psi\psi\, (25)} &=& {\Gamma_{(1)}^{r}}_{\alpha}^{~\gamma} {\Gamma_{(3)}^{p q r}}_{\delta}^{~\beta} \delta^{ik} \delta^{jl} ~~~,~~~ 
{\cal P}^{\QEDs}_{\psi\psi\, (26)} ~=~ {\Gamma_{(1)}^{r}}_{\delta}^{~\beta} {\Gamma_{(3)}^{p q r}}_{\alpha}^{~\gamma} \delta^{ik} \delta^{jl} 
\nonumber \\
{\cal P}^{\QEDs}_{\psi\psi\, (27)} &=& {\Gamma_{(1)}^{\mu}}_{\alpha}^{~\beta} {\Gamma_{(3)}^{p q \mu}}_{\delta}^{~\gamma} \delta^{ij} \delta^{kl} ~~~,~~~ 
{\cal P}^{\QEDs}_{\psi\psi\, (28)} ~=~ {\Gamma_{(1)}^{\mu}}_{\alpha}^{~\gamma} {\Gamma_{(3)}^{q r \mu}}_{\delta}^{~\beta} \delta^{ik} \delta^{jl} 
\nonumber \\
{\cal P}^{\QEDs}_{\psi\psi\, (29)} &=& {\Gamma_{(1)}^{\mu}}_{\delta}^{~\beta} {\Gamma_{(3)}^{p q \mu}}_{\alpha}^{~\gamma} \delta^{ik} \delta^{jl} ~~~,~~~ 
{\cal P}^{\QEDs}_{\psi\psi\, (30)} ~=~ {\Gamma_{(1)}^{\mu}}_{\delta}^{~\beta} {\Gamma_{(3)}^{p r \mu}}_{\alpha}^{~\gamma} \delta^{ik} \delta^{jl} 
\nonumber \\
{\cal P}^{\QEDs}_{\psi\psi\, (31)} &=& {\Gamma_{(1)}^{\mu}}_{\delta}^{~\gamma} {\Gamma_{(3)}^{p r \mu}}_{\alpha}^{~\beta} \delta^{ij} \delta^{kl} ~~~,~~~ 
{\cal P}^{\QEDs}_{\psi\psi\, (32)} ~=~ {\Gamma_{(1)}^{\mu}}_{\delta}^{~\gamma} {\Gamma_{(3)}^{q r \mu}}_{\alpha}^{~\beta} \delta^{ij} \delta^{kl} 
\nonumber \\
{\cal P}^{\QEDs}_{\psi\psi\, (33)} &=& {\Gamma_{(3)}^{p q r}}_{\alpha}^{~\beta} {\Gamma_{(3)}^{p q r}}_{\delta}^{~\gamma} \delta^{ij} \delta^{kl} ~~~,~~~ 
{\cal P}^{\QEDs}_{\psi\psi\, (34)} ~=~ {\Gamma_{(3)}^{p q r}}_{\alpha}^{~\gamma} {\Gamma_{(3)}^{p q r}}_{\delta}^{~\beta} \delta^{ik} \delta^{jl} 
\nonumber \\
{\cal P}^{\QEDs}_{\psi\psi\, (35)} &=& {\Gamma_{(3)}^{p q \sigma}}_{\alpha}^{~\beta} {\Gamma_{(3)}^{p q \sigma}}_{\delta}^{~\gamma} \delta^{ij} \delta^{kl} ~~~,~~~ 
{\cal P}^{\QEDs}_{\psi\psi\, (36)} ~=~ {\Gamma_{(3)}^{p q \sigma}}_{\alpha}^{~\beta} {\Gamma_{(3)}^{p r \sigma}}_{\delta}^{~\gamma} \delta^{ij} \delta^{kl} 
\nonumber \\
{\cal P}^{\QEDs}_{\psi\psi\, (37)} &=& {\Gamma_{(3)}^{p q \sigma}}_{\alpha}^{~\beta} {\Gamma_{(3)}^{q r \sigma}}_{\delta}^{~\gamma} \delta^{ij} \delta^{kl} ~~~,~~~ 
{\cal P}^{\QEDs}_{\psi\psi\, (38)} ~=~ {\Gamma_{(3)}^{p q \sigma}}_{\alpha}^{~\gamma} {\Gamma_{(3)}^{p q \sigma}}_{\delta}^{~\beta} \delta^{ik} \delta^{jl} 
\nonumber \\
{\cal P}^{\QEDs}_{\psi\psi\, (39)} &=& {\Gamma_{(3)}^{p q \sigma}}_{\alpha}^{~\gamma} {\Gamma_{(3)}^{p r \sigma}}_{\delta}^{~\beta} \delta^{ik} \delta^{jl} ~~~,~~~ 
{\cal P}^{\QEDs}_{\psi\psi\, (40)} ~=~ {\Gamma_{(3)}^{p q \sigma}}_{\alpha}^{~\gamma} {\Gamma_{(3)}^{q r \sigma}}_{\delta}^{~\beta} \delta^{ik} \delta^{jl} 
\nonumber \\
{\cal P}^{\QEDs}_{\psi\psi\, (41)} &=& {\Gamma_{(3)}^{p q \sigma}}_{\delta}^{~\beta} {\Gamma_{(3)}^{p r \sigma}}_{\alpha}^{~\gamma} \delta^{ik} \delta^{jl} ~~~,~~~ 
{\cal P}^{\QEDs}_{\psi\psi\, (42)} ~=~ {\Gamma_{(3)}^{p q \sigma}}_{\delta}^{~\beta} {\Gamma_{(3)}^{q r \sigma}}_{\alpha}^{~\gamma} \delta^{ik} \delta^{jl} 
\nonumber \\
{\cal P}^{\QEDs}_{\psi\psi\, (43)} &=& {\Gamma_{(3)}^{p q \sigma}}_{\delta}^{~\gamma} {\Gamma_{(3)}^{p r \sigma}}_{\alpha}^{~\beta} \delta^{ij} \delta^{kl} ~~~,~~~ 
{\cal P}^{\QEDs}_{\psi\psi\, (44)} ~=~ {\Gamma_{(3)}^{p q \sigma}}_{\delta}^{~\gamma} {\Gamma_{(3)}^{q r \sigma}}_{\alpha}^{~\beta} \delta^{ij} \delta^{kl} 
\nonumber \\
{\cal P}^{\QEDs}_{\psi\psi\, (45)} &=& {\Gamma_{(3)}^{p r \sigma}}_{\alpha}^{~\beta} {\Gamma_{(3)}^{p r \sigma}}_{\delta}^{~\gamma} \delta^{ij} \delta^{kl} ~~~,~~~ 
{\cal P}^{\QEDs}_{\psi\psi\, (46)} ~=~ {\Gamma_{(3)}^{p r \sigma}}_{\alpha}^{~\beta} {\Gamma_{(3)}^{q r \sigma}}_{\delta}^{~\gamma} \delta^{ij} \delta^{kl} 
\nonumber \\
{\cal P}^{\QEDs}_{\psi\psi\, (47)} &=& {\Gamma_{(3)}^{p r \sigma}}_{\alpha}^{~\gamma} {\Gamma_{(3)}^{p r \sigma}}_{\delta}^{~\beta} \delta^{ik} \delta^{jl} ~~~,~~~ 
{\cal P}^{\QEDs}_{\psi\psi\, (48)} ~=~ {\Gamma_{(3)}^{p r \sigma}}_{\alpha}^{~\gamma} {\Gamma_{(3)}^{q r \sigma}}_{\delta}^{~\beta} \delta^{ik} \delta^{jl} 
\nonumber \\
{\cal P}^{\QEDs}_{\psi\psi\, (49)} &=& {\Gamma_{(3)}^{p r \sigma}}_{\delta}^{~\beta} {\Gamma_{(3)}^{q r \sigma}}_{\alpha}^{~\gamma} \delta^{ik} \delta^{jl} ~~~,~~~ 
{\cal P}^{\QEDs}_{\psi\psi\, (50)} ~=~ {\Gamma_{(3)}^{p r \sigma}}_{\delta}^{~\gamma} {\Gamma_{(3)}^{q r \sigma}}_{\alpha}^{~\beta} \delta^{ij} \delta^{kl} 
\nonumber \\
{\cal P}^{\QEDs}_{\psi\psi\, (51)} &=& {\Gamma_{(3)}^{p \nu \sigma}}_{\alpha}^{~\beta} {\Gamma_{(3)}^{p \nu \sigma}}_{\delta}^{~\gamma} \delta^{ij} \delta^{kl} ~~~,~~~ 
{\cal P}^{\QEDs}_{\psi\psi\, (52)} ~=~ {\Gamma_{(3)}^{p \nu \sigma}}_{\alpha}^{~\beta} {\Gamma_{(3)}^{q \nu \sigma}}_{\delta}^{~\gamma} \delta^{ij} \delta^{kl} 
\nonumber \\
{\cal P}^{\QEDs}_{\psi\psi\, (53)} &=& {\Gamma_{(3)}^{p \nu \sigma}}_{\alpha}^{~\beta} {\Gamma_{(3)}^{r \nu \sigma}}_{\delta}^{~\gamma} \delta^{ij} \delta^{kl} ~~~,~~~ 
{\cal P}^{\QEDs}_{\psi\psi\, (54)} ~=~ {\Gamma_{(3)}^{p \nu \sigma}}_{\alpha}^{~\gamma} {\Gamma_{(3)}^{p \nu \sigma}}_{\delta}^{~\beta} \delta^{ik} \delta^{jl} 
\nonumber \\
{\cal P}^{\QEDs}_{\psi\psi\, (55)} &=& {\Gamma_{(3)}^{p \nu \sigma}}_{\alpha}^{~\gamma} {\Gamma_{(3)}^{q \nu \sigma}}_{\delta}^{~\beta} \delta^{ik} \delta^{jl} ~~~,~~~ 
{\cal P}^{\QEDs}_{\psi\psi\, (56)} ~=~ {\Gamma_{(3)}^{p \nu \sigma}}_{\alpha}^{~\gamma} {\Gamma_{(3)}^{r \nu \sigma}}_{\delta}^{~\beta} \delta^{ik} \delta^{jl} 
\nonumber \\
{\cal P}^{\QEDs}_{\psi\psi\, (57)} &=& {\Gamma_{(3)}^{p \nu \sigma}}_{\delta}^{~\beta} {\Gamma_{(3)}^{q \nu \sigma}}_{\alpha}^{~\gamma} \delta^{ik} \delta^{jl} ~~~,~~~ 
{\cal P}^{\QEDs}_{\psi\psi\, (58)} ~=~ {\Gamma_{(3)}^{p \nu \sigma}}_{\delta}^{~\beta} {\Gamma_{(3)}^{r \nu \sigma}}_{\alpha}^{~\gamma} \delta^{ik} \delta^{jl} 
\nonumber \\
{\cal P}^{\QEDs}_{\psi\psi\, (59)} &=& {\Gamma_{(3)}^{p \nu \sigma}}_{\delta}^{~\gamma} {\Gamma_{(3)}^{q \nu \sigma}}_{\alpha}^{~\beta} \delta^{ij} \delta^{kl} ~~~,~~~ 
{\cal P}^{\QEDs}_{\psi\psi\, (60)} ~=~ {\Gamma_{(3)}^{p \nu \sigma}}_{\delta}^{~\gamma} {\Gamma_{(3)}^{r \nu \sigma}}_{\alpha}^{~\beta} \delta^{ij} \delta^{kl} 
\nonumber \\
{\cal P}^{\QEDs}_{\psi\psi\, (61)} &=& {\Gamma_{(3)}^{q r \sigma}}_{\alpha}^{~\beta} {\Gamma_{(3)}^{q r \sigma}}_{\delta}^{~\gamma} \delta^{ij} \delta^{kl} ~~~,~~~ 
{\cal P}^{\QEDs}_{\psi\psi\, (62)} ~=~ {\Gamma_{(3)}^{q r \sigma}}_{\alpha}^{~\gamma} {\Gamma_{(3)}^{q r \sigma}}_{\delta}^{~\beta} \delta^{ik} \delta^{jl} 
\nonumber \\
{\cal P}^{\QEDs}_{\psi\psi\, (63)} &=& {\Gamma_{(3)}^{q \nu \sigma}}_{\alpha}^{~\beta} {\Gamma_{(3)}^{q \nu \sigma}}_{\delta}^{~\gamma} \delta^{ij} \delta^{kl} ~~~,~~~ 
{\cal P}^{\QEDs}_{\psi\psi\, (64)} ~=~ {\Gamma_{(3)}^{q \nu \sigma}}_{\alpha}^{~\beta} {\Gamma_{(3)}^{r \nu \sigma}}_{\delta}^{~\gamma} \delta^{ij} \delta^{kl} 
\nonumber \\
{\cal P}^{\QEDs}_{\psi\psi\, (65)} &=& {\Gamma_{(3)}^{q \nu \sigma}}_{\alpha}^{~\gamma} {\Gamma_{(3)}^{q \nu \sigma}}_{\delta}^{~\beta} \delta^{ik} \delta^{jl} ~~~,~~~ 
{\cal P}^{\QEDs}_{\psi\psi\, (66)} ~=~ {\Gamma_{(3)}^{q \nu \sigma}}_{\alpha}^{~\gamma} {\Gamma_{(3)}^{r \nu \sigma}}_{\delta}^{~\beta} \delta^{ik} \delta^{jl} 
\nonumber \\
{\cal P}^{\QEDs}_{\psi\psi\, (67)} &=& {\Gamma_{(3)}^{q \nu \sigma}}_{\delta}^{~\beta} {\Gamma_{(3)}^{r \nu \sigma}}_{\alpha}^{~\gamma} \delta^{ik} \delta^{jl} ~~~,~~~ 
{\cal P}^{\QEDs}_{\psi\psi\, (68)} ~=~ {\Gamma_{(3)}^{q \nu \sigma}}_{\delta}^{~\gamma} {\Gamma_{(3)}^{r \nu \sigma}}_{\alpha}^{~\beta} \delta^{ij} \delta^{kl} 
\nonumber \\
{\cal P}^{\QEDs}_{\psi\psi\, (69)} &=& {\Gamma_{(3)}^{r \nu \sigma}}_{\alpha}^{~\beta} {\Gamma_{(3)}^{r \nu \sigma}}_{\delta}^{~\gamma} \delta^{ij} \delta^{kl} ~~~,~~~ 
{\cal P}^{\QEDs}_{\psi\psi\, (70)} ~=~ {\Gamma_{(3)}^{r \nu \sigma}}_{\alpha}^{~\gamma} {\Gamma_{(3)}^{r \nu \sigma}}_{\delta}^{~\beta} \delta^{ik} \delta^{jl} 
\nonumber \\
{\cal P}^{\QEDs}_{\psi\psi\, (71)} &=& {\Gamma_{(3)}^{\mu \nu \sigma}}_{\alpha}^{~\beta} {\Gamma_{(3)}^{\mu \nu \sigma}}_{\delta}^{~\gamma} \delta^{ij} \delta^{kl} ~~~,~~~ 
{\cal P}^{\QEDs}_{\psi\psi\, (72)} ~=~ {\Gamma_{(3)}^{\mu \nu \sigma}}_{\alpha}^{~\gamma} {\Gamma_{(3)}^{\mu \nu \sigma}}_{\delta}^{~\beta} \delta^{ik} \delta^{jl} ~. 
\end{eqnarray}

Finally, we give the projection matrix for the ghost-quark $4$-point function
as an aid to orient with the information presented in the associated data file.
It is the most compact of the projections we constructed. We have 
\begin{equation}
{\cal M}_{c\psi} ~=~ \frac{1}{64} \left(
\begin{array}{cccc}
-~ 24 & -~ 12 & -~ 12 & 0 \\ 
-~ 12 & -~ 24 & -~ 12 & 0 \\ 
-~ 12 & -~ 12 & -~ 24 & 0 \\ 
0 & 0 & 0 & 27 \\
\end{array}
\right) ~.
\end{equation} 
The entries in this projection matrix are numbers in contrast with the others
we constructed which involve rational polynomials in $d$. This is because there
are no Lorentz indices on the elements of the ghost-gluon Lorentz tensor basis.

\end{document}